\documentclass[12pt]{article}
\usepackage{color}
\usepackage{amsmath, amssymb}
\usepackage{epsfig, palatino}
\usepackage{pstricks,pst-node,pst-tree}
\usepackage{epic}

\usepackage{mathrsfs}

\usepackage{ae} % or {zefonts}0
\usepackage[T1]{fontenc}
\usepackage[ansinew]{inputenc}
\usepackage{amsmath}
\usepackage{amssymb}
\usepackage{graphicx}
\usepackage{color}
\definecolor{darkblue}{cmyk}{0.9,0.9,0,0}
\usepackage[colorlinks=true,linkcolor=darkblue,citecolor=darkblue,urlcolor=darkblue]{hyperref}
\newcommand{\beq}{\begin{equation}}
\newcommand{\eeq}{\end{equation}}
\newcommand\beqa{\begin{eqnarray}}
\newcommand\eeqa{\end{eqnarray}}
\newcommand\bea{\begin{array}}
\newcommand\eea{\end{array}}

\def\XXint#1#2#3{{\setbox0=\hbox{$#1{#2#3}{\int}$}
\vcenter{\hbox{$#2#3$}}\kern-.5\wd0}}

\newcommand{\nn}{\nonumber}

\newcommand{\COMMENT}[1]{}

\newcommand{\neqa}{\nonumber\end{eqnarray}}
\newcommand{\la}[1]{\label{#1}}

\renewcommand{\d}{\partial}

\newcommand{\<}{{\langle}}
\renewcommand{\>}{{\rangle}}

\newcommand{\re}{\relax{\rm I\kern-.18em R}}

\def\su2{{SU(2)}}

\def\[{\left[}
\def\]{\right]}

\def\({\left(}
\def\){\right)}
\def\[{\left[}
\def\]{\right]}

\def\<{\langle}
\def\>{\rangle}

\def\i2{\frac{i}{2}}

\def\ds{\displaystyle}

 \def\nref#1{{(\ref{#1})}}

\usepackage{varioref}
\usepackage{makeidx}
\makeindex
\usepackage[english]{babel}

\begin{document}

%%%%%%%%%%%%%%%%%%%%%%%%%%%%%%%%%%%%%
%%%%%%%%%%%%%%%%%%%%%%%%%%%%%%%%%%%%%
%%%%%%%%%%%%%%%%%%%%%%%%%%%%%%%%%%%%%
%

\thispagestyle{empty}
%\begin{flushright}\footnotesize
%\texttt{arxiv:yymm.nnnn}\\
%\texttt{AEI-2009-078}\\
%\vspace{1.7cm}
%\end{flushright}

\renewcommand{\thefootnote}{\fnsymbol{footnote}}
\setcounter{footnote}{0}
\setcounter{figure}{0}
\begin{center}
$$$$
{\Large\textbf{\mathversion{bold} Bootstrapping Null Polygon Wilson Loops}\par}

\vspace{1.0cm}

\textrm{Davide Gaiotto${}^a$, Juan Maldacena$^a$, Amit Sever$^b$, Pedro Vieira$^b$}
\\ \vspace{1.2cm}
\footnotesize{

\textit{$^{a}$ School of Natural Sciences,\\Institute for Advanced Study, Princeton, NJ 08540, USA.} \\
\texttt{} \\
\vspace{3mm}
\textit{$^b$
Perimeter Institute for Theoretical Physics\\ Waterloo,
Ontario N2J 2W9, Canada} \\
\texttt{}
\vspace{3mm}
}

%%%%%%%%

\par\vspace{1.5cm}

\textbf{Abstract}\vspace{2mm}
\end{center}

\noindent
%\small

We derive the two loop expressions for polygonal Wilson loops by starting from the one loop expressions and applying an operator product expansion.
We do this for polygonal Wilson loops in ${R}^{1,1}$ and find a result in agreement with previous computations. We also discuss the spectrum of excitations around flux tube that connects two null Wilson lines.

\vspace*{\fill}

\setcounter{page}{1}
\renewcommand{\thefootnote}{\arabic{footnote}}
\setcounter{footnote}{0}

\newpage
\tableofcontents

\section{Introduction}

In this article we provide a short derivation for two loop  
polygon Wilson loop expectation value (or MHV amplitudes) using the
Operator Product Expansion for Wilson loops introduced in \cite{OPEpaper}.
This is done in a restricted kinematical region where the Wilson loop
is embedded in an ${R}^{1,1}$ subspace of four dimensional Minkowski space.

The idea is very simple.  One starts from the one loop result and then performs the OPE expansion. Each term of this  expansion corresponds
to  the exchange of a free particle. Next we include the one loop energy or anomalous dimension for each particle.
%This gives us the discontinuity of the two loop answer around the factorization locus.
The expansion breaks the cyclic symmetry of the answer.
If one completes it in the simplest way one obtains a two loop answer that has the correct expansion in all the OPE channels.
The computation is particularly simple because at two loops one can organize the
states into $SL(2,R)$ representations. With  ${R}^{1,1}$ kinematics only one $SL(2)$ representation
contributes (for each momentum), hence all states have the same anomalous dimension.

The two loop functions obtained in this fashion agree with explicit results
computed by more direct methods \cite{OctagonDuca} \cite{Anastasiou:2009kna} or by other
inspired guesswork \cite{Heslop:2010kq}. These were checked numerically in \cite{Heslop:2010kq} against
the direct perturbative computation.

This paper is organized as follows. We first review the OPE expansion for Wilson loops and explain
why we have an $SL(2)$ symmetry at two loops. We then discuss the derivation of the two loop results
for the octagon and decagon from this points of view, leaving the general $n$ case for appendix \ref{AnyN}.
We then say a few words about the three loop octagon and present some conclusions.
In appendix \ref{IntegrabilityAppendix} we discuss some aspects of the spectrum of excitations around the
flux tube that stretches between two null Wilson lines. This can also be viewed as the spectrum of excitations
around the infinite spin limit of finite twist operators, or the GKP \cite{Gubser:2002tv}
string.

\section{OPE review}\la{rev}

The Operator Product Expansion for  polygonal null Wilson loops is a certain expansion for the
expectation value of a Wilson loop correlator   in the limit where two or more consecutive edges become collinear. For more details see \cite{OPEpaper}.

\begin{figure}[t]
\begin{center}
 \includegraphics[width=.9\linewidth]{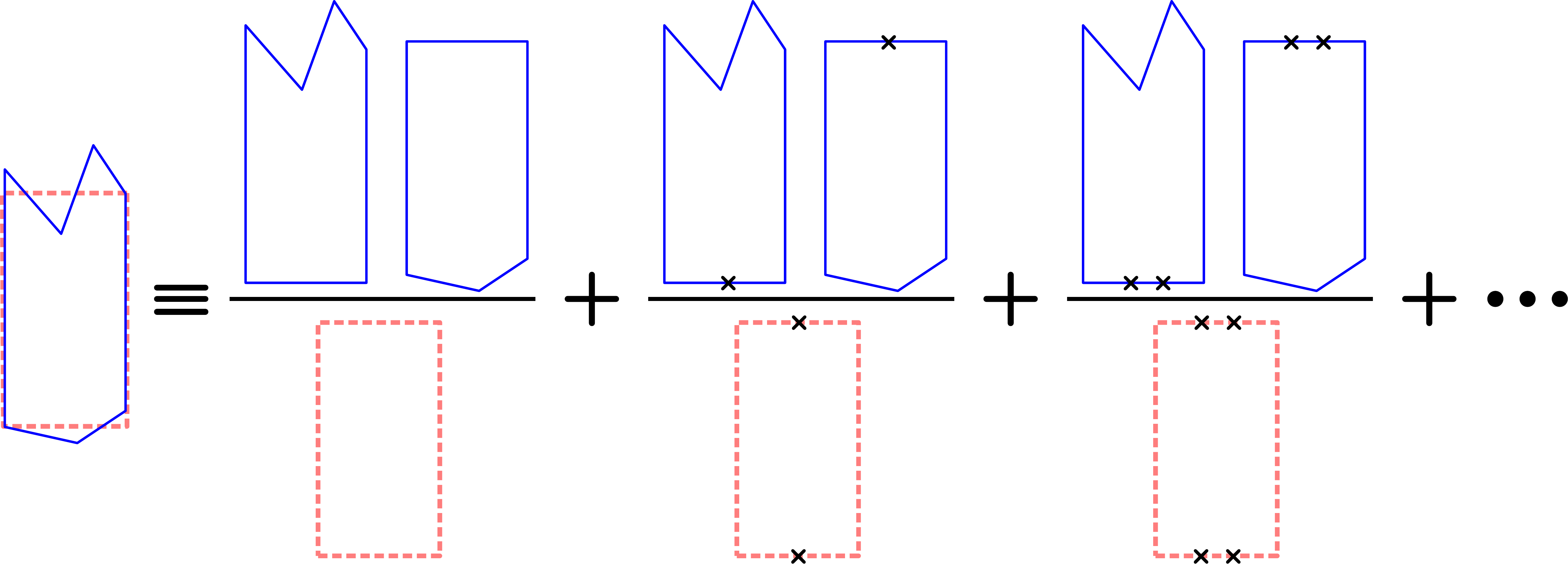}
\end{center}
\caption{We start from a general Wilson loop. We select two null lines, denoted in red. We then act with symmetries that are preserved by these null lines on one of the sides. The two null lines preserve a certain $SL(2)$ symmetry. Selecting a dilatation operator inside this $SL(2)$ amounts to a choice of a reference square, denoted here by the
red dashed lines.   This allows us to perform an expansion which has the rough form seen on the right. The first term comes from the exchange of the flux tube vacuum. The second from the exchange of a single excitation on the flux tube, the third from two excitations, etc. The denominators can be viewed as normalization factors, as we usually have in the standard OPE. }\label{OPEexpansion}
\end{figure}

We consider a polygonal Wilson loop made out of $n$ null segments.
  The collinear limit is approached as follows. We choose
 two null lines that belong to the Wilson loop polygon. These null lines are
then extended. The two null lines preserve an $SL(2) \times R_\sigma \times SO(2)_{\phi}$ subgroup of the conformal group. Using these null lines and two other arbitrary null lines we define a reference ``square''. This selects a
dilatation operator $R_\tau$ inside $SL(2)$, see figure \ref{OPEexpansion}. This reference square is invariant under three commuting symmetries $R_\tau \times R_\sigma \times SO(2)_{\phi}$.
Two of the symmetries are non-compact and are particular elements of the conformal group.
Then a family of polygons is constructed by acting on the bottom  part of the original polygon with these symmetries and
joining it to the original top  part of the polygon.
Schematically we have
\beq \la{actre}
 \langle W \rangle (\tau,\sigma,\phi) \equiv \langle {\rm top } | e^{ - \tau E + i \sigma P + i \phi J } | {\rm bottom} \rangle
 \eeq
 This has an expansion in terms of intermediate states which are excitations of the flux tube that goes between the
 two selected null sides. Such states can also be described in terms of excitations around the infinite spin limit of
 local operators with spin. This correspondence with local operators enables the computation of the properties of
  the propagating states.

 A slight complication is that, strictly speaking, the Wilson loop correlation is zero. That is due to certain well understood UV divergencies. After introducing a UV cut-off one remains with a finite answer. These
 symmetries are broken in a precise way determined by an explicitly known anomaly \cite{Drummond:2007aua}. We can thus take into account
 the effects of the anomaly.

 A standard way to take it into account is to consider the ratio function
 \beq \la{ratiof}
  e^R = { \langle W \rangle \over \left[ \langle W \rangle_{U(1)} \right]^{\Gamma_{cusp} } }
  \eeq
  where the denominator is the same Wilson loop but in a $U(1)$ theory, with the coupling replaced by $\Gamma_{cusp}$.  This ratio is a finite quantity, free of anomaly, known as the \textit{remainder function} \cite{Drummond:2007aua}.   Unfortunately, in the OPE expansion of $R$ one has already subtracted off some interesting contributions from the OPE expansion of  $\langle W \rangle_{U(1)} $, which is non-trivial. This makes it difficult to apply our program to directly to $R$.

\begin{figure}[t]
\begin{center}
 \includegraphics[width=.5\linewidth]{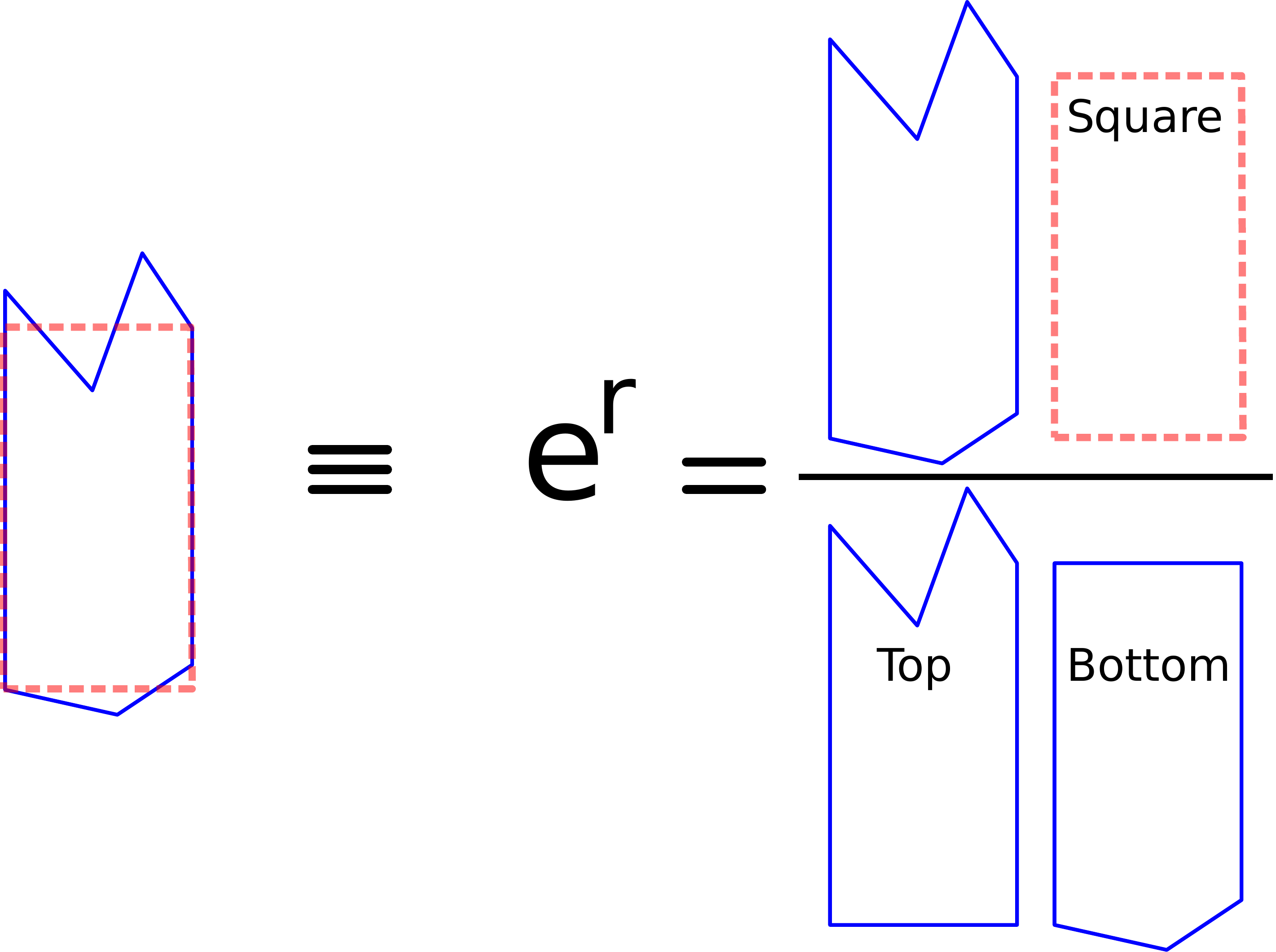}
\end{center}
\caption{ Definition of a ratio function that is finite and conformal invariant.
It involves the selection of a reference square whose vertices (across a diagonal) coincide with vertices of the original polygon. }\label{RatioWell}
\end{figure}

For us, in order to perform the OPE, it is convenient to introduce the following ratio function
  \beq\la{rbds}
r  =\log\(\frac{\< W\> \< W^{\rm square} \> }{\< W^{\rm top}\> \< W^{\rm bottom}\> }\)
\eeq
where the ``square'', ``top" and ``bottom" polygons are defined in figure \ref{RatioWell}.
 We can get rid of the divergencies completely
 if we choose a reference square that
coincides with some of the cusps of the original polygon, see figure \ref{RatioWell}.\footnote{
For a general reference square,  the ratio \nref{rbds} does not quite get rid of divergencies,
 it leaves some unimportant  single logarigthmic divergencies which
have zero momentum (under $R_\sigma$). Here we will make the choice in figure \ref{RatioWell} which gets
rid of this ambiguity.} In this
fashion we can have a well defined and non-vanishing answer already at one loop.  The price we pay is however that $r$ is not cyclic invariant.  The remainder function differs from $r$ in a simple way \cite{OPEpaper}
\beq
R-r=R_{\text{top}}+R_{\text{bottom}}-\Gamma_{cusp}  r_{U(1)} \,.
\eeq
and only the last term contributes to the OPE expansion.

As an example consider the octagon in $R^{1,1}$ kinematics, see figure \ref{OctagonFig}.a. We label the location of the cusps as
\beq\la{points}
\{\dots,(x_i^+,x_{i-1}^-),(x_i^+,x_{i}^-),(x_{i+1}^+,x_{i}^-),\dots\}\ ,
\eeq
where $x^\pm=x\pm t$. For the Octagon, we fix these locations at
\beqa
x^+_i=\{-\infty,-1,0,e^{2\tau}\}\ ,\qquad x^-_{i}=\{-\infty,-1,0,e^{-2\sigma}\}\ .
\eeqa
The corresponding  ratio function at one loop  reads
\beq
r_{U(1)}^{\,octagon}=-{g^2 \over2}\log\(1+e^{-2\tau}\)\log\(1+e^{-2\sigma}\) \la{octagon}
\eeq
This is the ``seed'' that we will use to get the two loop answer. This one loop result is the same as the
result we would obtain in a $U(1)$ theory and it only comes from a single gluon (or single photon) exchange
between the various null segments.

\subsection{$SL(2,R)$ symmetry at two loops }

When we choose two null lines we preserve an $SL(2) \times R_\sigma \times SO(2)_{\phi}$ symmetry which is a
subgroup of the full conformal group, $SO(2,4)$.
Here we will make some remarks about the $SL(2)$ symmetry. This $SL(2)$ symmetry is broken by the
flux tube. However, the mere existence of a flux tube is an effect of order $g^2$. In fact, we will
see that some consequences of the $SL(2)$ symmetry are still preserved at low enough orders in perturbation
theory.
Let us start with the one loop answer. This arises from the propagation of free particles between the edges
of a square, as in figure \ref{OPEexpansion}. A factor of $g^2$ already rises by creating and 
annihilating the excitation from
the Wilson contour. Thus, the propagation in the bulk occurs as in the free theory. In the free theory this
$SL(2)$ is, of course, a symmetry, since it is just a subgroup of the full conformal group. Thus, the particles
being exchanged form multiplets under this symmetry.  An $SL(2)$ representation consists
of a primary $P$, or lowest weight state, and an infinite set of descendents of the form $P_n = (L_{-1})^n P $.
More explicitly, in the case of $R^{1,1}$ kinematics, the only relevant operator is $F_{+-}$ and
the descendents are $D_-^n F_{+-}$. The twist of these operators is  $ \epsilon_0 = 2 +  2 n$.
The $D_+$ derivatives are essentially
taken into account by the momentum quantum number, related to the $R_\sigma$.
\begin{figure}[t]
\begin{center}
\includegraphics[width=.6\linewidth]{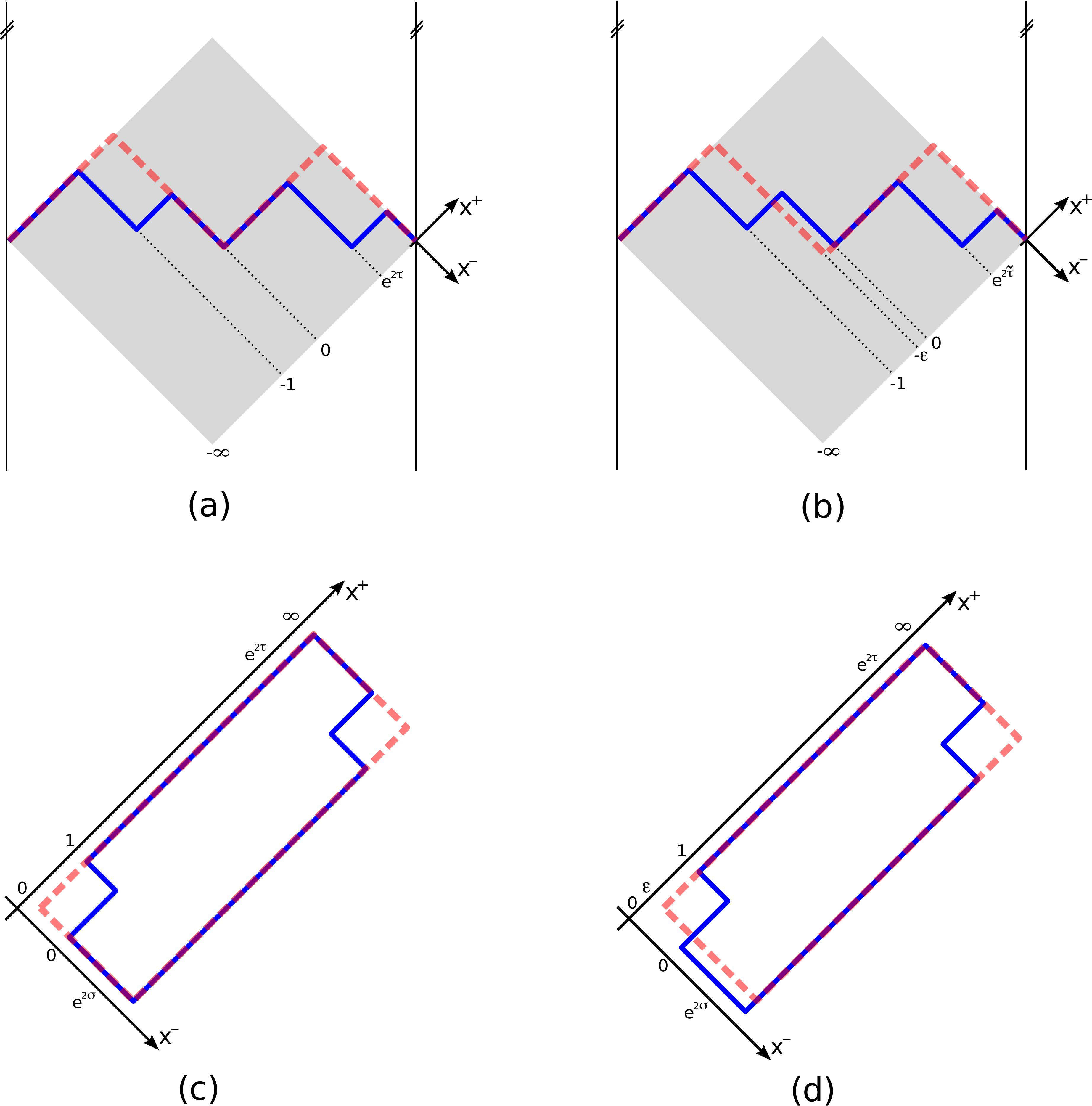}
\end{center}
\caption{(a) The Octagon null Wilson loop in blue embedded in the Penrose diagram of $R^{1,1}$. The red dashed line is the reference square we start with. (b) A different choice of reference square that is suitable for an OPE expansion in the same channel as in (a). In that example, the two choices of reference squares only differ by the position of the "bottom" cusp at 0 (a) and $-\epsilon$ (b).  The corresponding OPE expansion parameters are related by the infinitesimal transformation $\tau\to\tau+{\epsilon \over 2} \,e^{-2\tau}$. We must be able to re-write the OPE expansion after this transformation in the same form as before with the same anomalous dimensions. That is only possible if $\gamma_k(p)$ is independent of $k$. In (c) and (d) we draw the same picture as it would have looked like if the Octagon under consideration was Lorentzian.}\label{OctagonFig}
\end{figure}

Now let us see what happens at two loops. At two loops, the particle starts feeling the effects of the color electric flux between the two null lines.
This flux  breaks the $SL(2)$ symmetry.
So, if we act with $L_{-1}$ on the flux vacuum, we can create
a particle, but only with amplitude $g$, since the background flux $F_{\mu \nu} \sim g^2$.
In addition, the anomalous dimensions of the $n$-th single particle
descendent, $P_n$ could be changed, to
$
\epsilon = 2 + 2 n + g^2 \gamma_n + o(g^4)$ .
It is possible to show that   $\gamma_n $ is still independent of $n$, so that
 we have
 \beq \label{nice}
 \epsilon = 2 + 2 n + g^2 \gamma(p) + o(g^4 ) ~,~~~~~~~~~n=0,1,2,\cdots
 \eeq
where we noted that $\gamma$ can
depend on $p$, which is the quantum number under the translation $R_\sigma$.
This can be shown by starting with a single particle state $P_n|0\rangle$. Acting with $L_{-1}$ we get
$ P_{n+1} |0 \rangle + g P_n P_0 | 0 \rangle$, where the second term is a two particle state arising from
the non-invariance of the vacuum.
 By acting with both sides of the commutator
$ L_{-1} = [L_0,L_{-1}]$ on $P_n |0\rangle$,
we can show that $\gamma_n$ should be independent of $n$.

It turns out that one can derive the same condition by demanding that the OPE expansion should have
the same form with different choices of reference square, see figure \ref{OctagonFig}.
 It should  have the same form, but with different
OPE coefficients.

\section{Two loop  polygonal Wilson loops in ${R}^{1,1}$}

\subsection{The Octagon}\la{Octagon}

In this section we   explain how to derive the octagon two loop remainder function up to an overall constant from the symmetries of the square together with the existence of the OPE limit. We start by writing down the OPE expansion of the $U(1)$ result (\ref{octagon}),
\beqa
r_{U(1)}^{\,octagon}=-{g^2 \over2}\log\(1+e^{-2\tau}\)\log\(1+e^{-2\sigma}\) = \sum_{k=1}^\infty \int dp \, C_k^{(1)}(p)  \,e^{ip\sigma-2k \tau}  \la{expU1} \,.
\eeqa
The expansion (\ref{expU1}) is a sum over twist $2k$ particles in the free theory where their energy is quantized and equal to $2k$. The constants $C_k^{(1)}$ represent the amplitude for creating and annihilating these particles on the top and bottom parts of the octagon. They are the one loop form factors or structure constants.
We could of course compute $C_k^{(1)}(p)$ by a simple Fourier transform\footnote{
We would get
$C_k^{(1)}(p)=g^2 (-1)^k/ (4 kp\sinh(\pi p/2))
$ and the integral over $p$ in (\ref{expU1}) should be understood as going slightly above the real axis.} however we will not need to know their precise expressions. It suffices to notice that the $k$ dependence is very simple,
 since we are expanding $\log(1+e^{-2\tau})$,\footnote{ The function $\log(1+e^{-2\tau})$ is nothing but the $SL(2)$ conformal block for a dimension $1$ primary. This is what we should expect given the discussion of the previous section. Therefore, to a certain extent we do not even need the $U(1)$ result for this argument.}
\beq
C_k^{(1)}(p)=\frac{(-1)^k}{k} C^{(1)}(p)
\eeq
We can now move to two loops. At two loops a few things change. On the one hand the single particle form factors get corrected. On the other hand the single particle energies get an anomalous dimension of order $g^2$. This second contribution gives a term linear in $\tau$ of the form
\beqa
\(r_{\,\text{2 loops}}^{\,\text{octagon}}\)_{\text{linear in }\tau}=- g^4 \tau \sum_{k=1}^\infty \frac{(-1)^k}{k} e^{-2k \tau}\int dp \, C^{(1)}(p)  \,e^{ip\sigma}   \gamma_k(p) \la{expU1lintau}
\eeqa
which comes from expanding the exponential  $e^{ - \tau \epsilon_k}$ in powers of $g^2$ and using \nref{nice}.
According to (\ref{nice}) the anomalous dimensions $\gamma_k(p)$ is actually independent of $k$ at this loop order. Hence we have
\beqa\la{linear}
\(r_{\,\text{2 loops}}^{\,\text{octagon}}\)_{\text{linear in }\tau}= -g^4 \tau \log(1+e^{-2\tau}) f(\sigma) \,. \la{linearInu}
\eeqa
 At this point, we should in principle compute the anomalous dimension $\gamma(p)$ and compute the function $f(\sigma)$.
There is an intuitive shortcut to the correct result.
 The remainder function $R$, is a cyclic invariant function and therefore $\tau \leftrightarrow -\tau$ symmetric. At two loop, its term linear in $\tau$ is given by (\ref{linear}). This allows us to  make the following simple
guess for the full $\tau$ dependence
\beq
R_{\,\text{2 loops}}^{\,\text{octagon}}=-{g^4\over2}\log(1+e^{2\tau})\log(1+e^{-2\tau})f(\sigma)
\eeq
Moreover, $R$ is parity invariant and therefore $\tau \leftrightarrow \sigma$ symmetric.
Now we use that symmetry to get $f(\sigma)$,
\beq
\la{voila} R_{\,\text{2 loops}}^{\,\text{octagon}}= -{g^4\over2}\log(1+e^{2\tau})\log(1+e^{-2\tau})\log(1+e^{2\sigma})\log(1+e^{-2\sigma})
\eeq
is agreement with \cite{OctagonDuca}.
The argument we presented, however,  does not fix the overall multiplicative constant in (\ref{voila}).

 To do that, we need an honest computation of $\gamma(p)$.  It is however easy to see that the normalization we picked is the correct one:
Fourier-transforming $f(\sigma)$ from (\ref{voila}) we can read the dimension of the twist $2k$ fields,
\beq
\epsilon_k(p)=2k+2g^2\[\psi(1+ip)+\psi(1-ip)-2\psi(1) \] \,, \la{dispersion}
\eeq
where $\psi(u)=(\log \Gamma)'(u)$.
The normalization of the $g^2$ term in the dispersion relation is directly connected to the normalization of (\ref{voila}). Since we can compute  \nref{dispersion} independently using $\mathcal{N}=4$ integrability we can fix the normalization of \nref{voila} \footnote{A shortcut is to recall that the large $p$ behavior of the one loop anomalous dimension of excitations around the GKP solution should diverge as $\Gamma\log(p) $
where $\Gamma$ is the cusp anomalous dimension which is known  \cite{Korchemsky:1995be,Belitsky:2003ys}. }, see appendix \ref{IntegrabilityAppendix} for details.

 It is rather simple to compute $f(\sigma)$ from the anomalous dimension \nref{dispersion}: the fact that the anomalous dimension is diagonal   in momentum space implies that in position
space it acts as a convolution kernel. By Fourier transforming \nref{dispersion} we find that the logarithmic
term of the two loop octagon is
\beqa
\(R_{\,\text{2 loops}}^{\,\text{octagon}}\)_{\text{linear in }\tau}&=& - \tau  g^2  \int_0^\infty dt
{ [ 2 R_1(\tau,\sigma) - R_1(\tau, \sigma + t/2) - R_1(\tau, \sigma - t/2 ) ] \over e^t -1 }
 \la{linearInusec}
\\
&=&  - { g^4 \over 2 }  \tau \log (1 + e^{- 2 \tau } ) \log(1 + e^{ - 2 \sigma } ) \log(1 + e^{ 2 \sigma } )
\eeqa
where $R_1$ is the one loop answer \nref{expU1}. Symmetrizing we get to \nref{voila}.

These arguments   allow  extra terms with an OPE expansion
 without linear terms at large (positive or negative) $\tau$ or  $\sigma$.
 A simple example of one such term is
\beq
\pi^2 \log\(1+\frac{1}{\cosh 2\tau}\)\log\(1+\frac{1}{\cosh 2\sigma}\) \,.
\eeq
This term respects the $\tau\to -\tau$, $\sigma \to -\sigma$ and $\tau \leftrightarrow \sigma$ symmetries of the square and has an OPE expansion without linear terms. Our previous argument would be blind to it, we would have to resort to a
 physical argument to discard such term.
In this particular case we could discard it since it has extra singularities when $\tau = \pm \pi/4 , \pm 3\pi/4$ etc, which should not be there.  It would be interesting to understand in more generality to which extent is $\mathcal{N}=4$ \textit{OPE constructible}.

In fact, the OPE is very similar to looking at particular physical cut contributions in amplitudes. It would
be interesting to find a precise relation.

\subsection{The Decagon}\la{Decagon}

In this section we use the same technique to derive the Decagon two loops remainder function. The generalization to any even $n>10$ is given in Appendix \ref{AnyN}. We start by gauge fixing the decagon cusps to be at
\begin{figure}[t]
\begin{center}
\includegraphics[width=.7\linewidth]{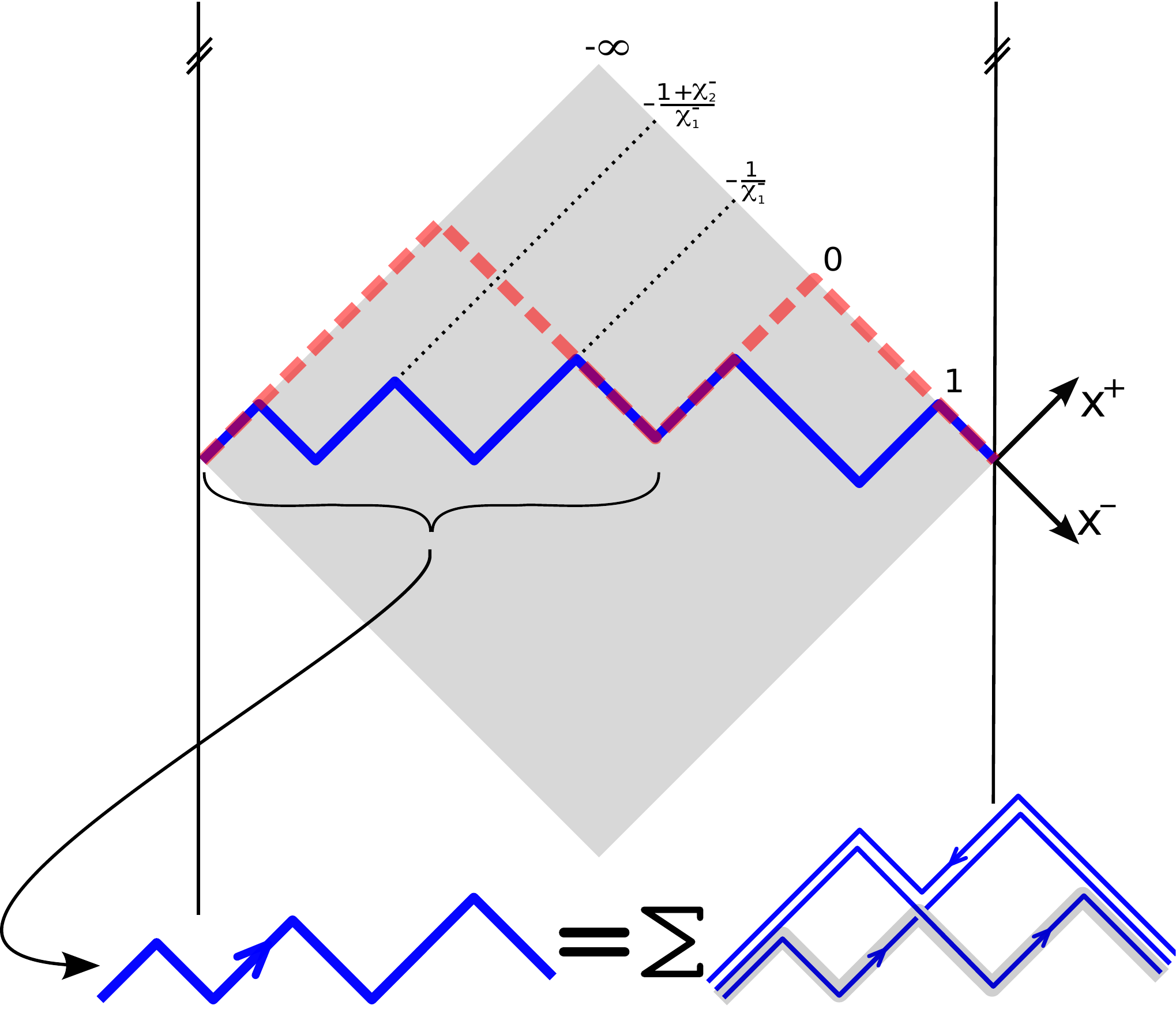}
\end{center}
\caption{The Decagon null Wilson loop in blue embedded in the Penrose diagram of $R^{1,1}$. The red dashed line is the reference square.}\label{DecagonFig}
\end{figure}
\beqa
&&x^+_i=\{-\infty,-\chi_1^+,-{\chi_1^+\over1+\chi_2^+},0,1\}\\
&&x^-_{i}=\{-\infty,-{1+\chi_2^-\over\chi_1^-},-{1\over\chi_1^-},0,1\}\ .\nn
\eeqa
The cusps are located as in (\ref{points}).
A corresponding basis of four conformal cross ratios is
\beq\la{DecCross}
\chi_1^+={x_{5,1}^+x_{4,2}^+\over x_{4,5}^+x_{1,2}^+}\ ,\qquad  \chi_2^+={x_{4,1}^+x_{3,2}^+\over x_{3,4}^+x_{1,2}^+}\ ,\qquad\chi_1^-={x_{1,3}^-x_{5,4}^-\over x_{5,1}^-x_{3,4}^-}\ \qquad \chi_2^-={x_{4,1}^-x_{3,2}^-\over x_{3,4}^-x_{1,2}^-}
\eeq
Any OPE limit corresponds to a choice of two non adjacent edges in the $x^+$ or $x^-$ directions. For the Decagon there are ten different such choices, i.e. ten different OPE limits. These limits are
 all related by the symmetries of the Decagon which reshuffle the cross ratios. These symmetries are
  {\it parity}, which reflects the sign of time, and {\it cyclicity},
   which cyclicly permutes the labeling of the $x^+$ and $x^-$ points. Combining these two we get a $\mathbb{Z}_{10}$ transformation that acts on the cusps points as
\beq\la{Z10}
x^-_i\to x^+_{i+1}\ ,\qquad x^+_i\to x^-_{i}\ .
\eeq
Under this transformation we have\footnote{The best way to find how that transformation acts on the conformal cross ratios (\ref{DecCross}) is  to map them to the $Y$ functions of \cite{Alday:2010vh}. The reason is that (\ref{Z10}) has a simple action on these cross ratios given by
$Y^{[a]}_s\to Y^{[a+1]}_s$. For the Decagon $Y_0=Y_3=0$ and $Y_s^{[a+10]}=Y_s^{[a]}$. The translation between the $Y$ functions and the cross ratios (\ref{DecCross}) can be read from \cite{Alday:2010vh} and is given by
\beq\la{Ymap}
\chi_1^+=Y_1^{[1]}=Y_2^{[6]},\quad \chi_2^+=Y_2=Y_1^{[5]},\quad \chi_1^-=Y_1=Y_2^{[5]},\quad \chi_2^-=Y_2^{[1]}=Y_1^{[6]}\ .
\eeq
We can then use the $Y$ system equations
$Y_s^{[a+1]}Y_s^{[a-1]}=(1+Y_{s+1}^{[a]})(1+Y_{s+1}^{[a]})$
to map all the $Y$ functions generated by the $\mathbb{Z}^{10}$ action to the ones appearing in (\ref{Ymap}). Using these equations one finds that (\ref{Z10}) acts on the cross ratios (\ref{DecCross}) as (\ref{Z10gen}). This procedure generalizes immediately for larger number of edges.}
\beq\la{Z10gen}
\chi_1^+\to{1+\chi_2^-\over\chi_1^-}\ ,\qquad\chi_1^-\to\chi_1^+\ ,\qquad\chi_2^+\to\chi_2^-\ ,\qquad\chi_2^-\to{1+\chi_1^+\over\chi_1^-}\ .
\eeq

We will now construct two loops Decagon OPE expansion in one channel. Demanding the answer to be $\mathbb{Z}_{10}$ symmetric will guaranty the correct OPE limit in all other channels.

{\bf The ${\bf U(1)}$ part}.
We choose the two null lines for the OPE expansion to be in the $x^{\pm}$ directions and located at $x^+=0$ and $x^+=\infty$. As before, we choose the reference square cusps on these lines to be located at $x^-=0$ and $x^-=\infty$ (see figure \ref{DecagonFig}). For that choice, the OPE expansion parameter is $\chi_1^-=e^{-2\tau}$, whereas $\chi_1^+=e^{-2\sigma}$. The $U(1)$ result in that channel is read from the one loop amplitude as explained in section \ref{rev}. It is given by
\beqa\la{U1part}
r_{U(1)}&=&-{g^2\over2}\log\[1+{\chi_1^-\over1+\chi_2^-}\]\log\[1+\chi_1^+\]-{g^2\over2}\log\[1+\chi_1^-\]\log\[1+{\chi_1^+\over1+\chi_2^+}\]\\&\ &+{g^2\over2}\log\[1+{\chi_1^-\over1+\chi_2^-}\]\log\[1+{\chi_1^+\over1+\chi_2^+}\]+[{\rm terms\ independent\ of\ \tau}] \ .\nn
\eeqa
To see that, note that the Decagon can be decomposed into three Octagons as in the bottom of figure \ref{DecagonFig}. Since the $U(1)$ part is linear, it decomposes into a sum over these three Octagons corresponding to the three terms in (\ref{U1part}). The same applies to any $n>8$ (see Appendix \ref{AnyN}).

{\bf The OPE procedure}.
 The expression (\ref{U1part}) is written as a sum over three terms, each of the form of the Octagon $U(1)$ part with shifted $\tau$ and $\sigma$. That is, each term has the form
\beqa
\log\(1+e^{-2\tau+a}\)\log\(1+e^{-2\sigma+b}\)
=\sum_{k=1}^\infty\(-e^{-2\tau+a}\)^k\frac{1}{k}\int{dp\ e^{-ip(\sigma+b/2)}\over 2(p-i0)\sinh(\pi p/2-i0)}
\eeqa
As for the Octagon, we identify the terms $e^{-2k\tau}$ in the sum with the exchange of the twist $2k$ one particle state. At two loops, that term will give rise to a term linear in $\tau$ obtained by dressing the previous expression by
\beq
-\tau\gamma_1(p)=-2\tau g^2\[\psi(1+ip/2)+\psi(1-ip/2)-2\psi(1)\]\ .
\eeq
We conclude that
\beqa\la{OPE}
\log\(1+e^{-2\tau+a}\)\log\(1+e^{-2\sigma+b}\) \xrightarrow[{\rm OPE}]{}2 g^2 \tau\log\(1+e^{-2\tau+a}\)\log\(1+e^{-2\sigma+b}\)\log\(1+e^{2\sigma-b}\)
\eeqa
By appling (\ref{OPE}) to the three terms in the $U(1)$ part (\ref{U1part}), we find the part of the two loop remainder function which is linear in $\tau$:
\beqa\la{expansion}
R^{(2)}_{\text linear\ in\ \tau}&=&\ {g^4\over2}\log\[\chi_1^-\]\log\[1+{\chi_1^-\over1+\chi_2^-}\]\log\[1+\chi_1^+\]\log\[1+{1\over\chi_1^+}\]\\&&-{g^4\over2}\log\[\chi_1^-\]\log\[1+{\chi_1^-\over1+\chi_2^-}\]\log\[1+{\chi_1^+\over1+\chi_2^+}\]\log\[1+{1+\chi_2^+\over\chi_1^+}\]\nn\\
&&+{g^4\over2}\log\[\chi_1^-\]\log\[1+\chi_1^-\]\log\[1+{\chi_1^+\over1+\chi_2^+}\]\log\[1+{1+\chi_2^+\over\chi_1^+}\]\nn
\eeqa

{\bf Symmetrization}.
This expression is not invariant under the $\mathbb {Z}_{10}$ symmetry and therefore cannot be the full result. We should now proceed like in the octagon case. I.e. we should first replace $\log(\chi_1^-)$ by some function which behaves as $\log(\chi_1^-)$ when $\chi_1^- \to 0$. Next we should symmetrize the result over the $\mathbb{Z}_{10}$ action. Now, in general, this symmetrization will spoil the OPE expansion (\ref{expansion}). The idea of the bootstrap program is to look for a replacement which preserves (\ref{expansion}).
Remarkably, the simple replacement $\log(\chi_1^-)\to -\log(1+1/\chi_1^-)$ does the job!
Of course, by construction, the remainder function obtained in this way is $\mathbb{Z}_{10}$ symmetric. We therefore arrive at the following expression for the two loop Decagon remainder function\footnote{This result was independently derived before \cite{Heslop:2010kq} appeared.}
\beqa
R_{\text 2\ loop}&=&-{g^4\over2}\log \[\frac{\chi_2^-}{\chi_1^-+1}+1\] \log \[\frac{\chi_1^-}{\chi_2^-+1}+1\] \log
   \[\frac{(\chi_1^++1) (\chi_2^++1)}{\chi_1^+ \chi_2^+}\] \log \[\frac{(\chi_1^++1)
   (\chi_2^++1)}{\chi_1^++\chi_2^++1}\]\nn\\
   &&-{g^4\over2}\log \[\frac{1}{\chi_2^-}+1\] \log \[\frac{(\chi_1^-+1)
   (\chi_2^-+1)}{\chi_1^-+\chi_2^-+1}\] \log \[\frac{\chi_1^++1}{\chi_2^+}+1\] \log
   \[\frac{\chi_2^+}{\chi_1^++1}+1\]\\
   &&-{g^4\over2}\log \[\frac{1}{\chi_1^-}+1\] \log
   \[\frac{(\chi_1^-+1) (\chi_2^-+1)}{\chi_1^-+\chi_2^-+1}\] \log \[\frac{\chi_1^++\chi_2^++1}{\chi_1^+}\]
   \log \[\frac{\chi_1^+}{\chi_2^++1}+1\]\nn\\
   &&-{g^4\over2}\log \[\frac{1}{\chi_1^-}+1\] \log
   \[\frac{1}{\chi_1^+}+1\] \log \[\chi_1^++1\] \log
   \[\frac{\chi_1^-}{\chi_2^-+1}+1\]\nn\\
   &&-{g^4\over2}\log \[\frac{1}{\chi_2^-}+1\] \log
   \[\frac{1}{\chi_2^+}+1\] \log \[\chi_2^++1\] \log \[\frac{\chi_1^-+\chi_2^-+1}{\chi_1^-+1}\] \ .\nn
\eeqa
which is compatible with the OPE expansion in all possible channels.
Equivalently,
\beqa
R_{\text 2\ loop}=-{g^4\over2}\log\[1+{1\over\chi_1^-}\]\log\[1+{\chi_1^-\over1+\chi_2^-}\]\log\[1+\chi_1^+\]\log\[1+{1\over\chi_1^+}\]+{\rm cyclic\ permutations}\ ,\nn
\eeqa
where cyclic permutation stands for the $\mathbb {Z}_5\subset\mathbb {Z}_{10}$ cyclic permutation of the $x^+$ and $x^-$ cusps. Up to a $-\pi^4/12$ constant, that is indeed the Decagon two loops remainder function guessed (and confirmed by numerics) in \cite{Heslop:2010kq}. In  Appendix \ref{AnyN} we generalize this result for $n>10$.

\subsection{ General gauge theories}

In the above results we have only used the symmetries of the problem.
Thus, we expect such results to be valid in any conformal gauge theory with a weak coupling limit.
It is also easy to understand from diagramatic point of view why
 the results are valid for any planar conformal gauge theory at two loops.
 Consider  all the two loop diagrams.
Many of the diagrams involve only gauge bosons and are the same in all theories. The
one diagram that is different is the bubble correction to the propagator. Such a diagram
contains a piece with gauge bosons and a piece that involves a  matter loop. The theory is a gauge
theory coupled to matter in such a way that it leads to a CFT with a tunable coupling,
with zero beta function (by assumption). The matter theory has an   global symmetry that
we are gauging. Let us call $j$ the corresponding currents. We have a current
two point function $\<jj\>\sim k/r^6$. The condition that the $\beta$ function vanishes is the condition that the
constant k has the value it has  for the matter in
$\mathcal{N} = 4$ SYM. Thus the diagrams involving the matter
bubble have the same value that they would have in
$\mathcal{N} = 4$ SYM.
This argument also appears to work for non-planar gauge theories (at two loops).
  In this case we have
different diagrams, but only involving gauge fields.

It would be interesting to investigate a similar argument
for the case of three dimensional conformal field theories, such as ABJM \cite{Aharony:2008ug}.

\section{ Remarks on the three loop octagon }

We can attempt to extend this method to the computation of the octagon Wilson loop at three loops,
still in $R^{1,1}$ kinematics. One difficulty is that at two loops we might be able to create propagating states
consisting of two particles.
Let us ignore this issue and let us first focus on one part of the answer which only involves the propagation of
one particle states.

At two loops we obtained the
 part of the two loop answer that goes like $\tau  F_2(\tau, \sigma)$, where
$F_2$ has an expansion in powers of $e^{ -  2 \tau }$.  This was computed by convolving
the anomalous dimension kernel with the one loop result, see \nref{linearInusec}.
At three loops we can easily obtain the piece that goes like
  $\tau^2 F_3(\tau, \sigma)$. We simply need to convolve twice with the anomalous
  dimension kernel appearing in \nref{linearInusec}. We
  obtain
\beqa
\tau^2 F_{3}(\tau, \sigma ) &= &  { (- \tau)^2 \over 2 }  \gamma * \gamma * R_1 = { \tau^2 \over 2 } ( 2 g^2 )^2 g^2  \log (1 + e^{- 2 \tau} ) f_3(\sigma)
\cr
f_3(\sigma ) & =&  - 2{\rm Li}_3(-e^{2\sigma})  +
 2 \sigma {\rm Li}_2(-e^{2\sigma}) - { 4 \over 3}
\log^3 (1 + e^{ 2 \sigma } ) +
\cr
 && + 4 \sigma \log^2(1 + e^{ 2 \sigma } ) - ( 2 \sigma^2 + { \pi^2 \over 6 } )  \log(1 + e^{ 2 \sigma} )   \label{threeexp}
\eeqa
Even though it is not manifest, we have that $f_3(\sigma) = f_3(-\sigma)$.
 If we now symmetrize this under $\tau \leftrightarrow \sigma $ we obtain
\beq \label{guessre}
R_3^{part} \propto  g^6 f_3(\tau) f_3(\sigma)
\eeq
As we take $\tau \to \infty$ we get the right $\tau^2 $ terms, in agreement with \nref{threeexp}.
One could wonder if \nref{guessre} is the full answer or not. If one   {\it assumes } that the
three loop answer factorizes into a function of $\tau $ and a function of $\sigma$, then
\nref{guessre} is the only consistent answer. In fact, if we assume that the answer factorizes at all loop
orders, then one can determine it using the same method. However, we do not know any reason why it should
factorize. And the strong coupling answer \cite{AMoct} appears  inconsistent with this factorization hypothesis.
 This does not say at what loop order
  factorization would stop.
There is another potential problem with \nref{guessre}. We can consider the term linear in $\tau$.
The guess \nref{guessre} gives a very specific term linear in $\tau$. On the other hand
the OPE analysis gives two sources for terms linear in $\tau$. We have one loop corrections to the OPE
coefficients
 and two loop corrections of the anomalous dimensions. The latter on its own gives a term which
has a rather different structure, being of transcendentality two in $\tau$ and four in $\sigma$. However, we
have not computed in detail the corrections to the OPE coefficients. This correction could involve the
creation of two particle states and we have not fully analyzed it. We leave this to the future.
 The conclusion is that probably \nref{guessre} is just a part of the three loop answer, which correctly captures the full $\tau^2$ behavior, but by some miraculous cancelations it could be the full answer.

\section{Conclusions and open problems}

 In this paper we have provided a quick derivation of the two loop result for polygonal Wilson loops in
 $R^{1,1}$ kinematics. It is based on an important feature of the OPE expansion of the two loop results:
the terms which have a logarithmic discontinuity around the factorization locus can be predicted from the OPE expansion of the one loop result.
 The idea is to start from the known result for the one loop Wilson loops. This is a simple function which
 is just the result that we would obtain in a $U(1)$ theory. We then apply the Operator Product Expansion
 for Wilson loops derived in \cite{OPEpaper}. One selects a particular expansion channel and the structure
 of the OPE, together with the knowledge of the anomalous dimensions, allows us to derive certain logarithmic
 terms in the two loop result. They are the discontinuity of the two loop answer around the factorization locus.
 One then guesses a simple way to complete these terms into a full expression.
 When we select an expansion channel we break the cyclic symmetry. Imposing the cyclic symmetry gives us
 a natural way to complete the answer into the full expression. In this way we rederived the two loop octagon
 expression \nref{voila}, derived originally in \cite{OctagonDuca}. We have extended these results for higher
 number of gluons. These expressions were guessed previously in \cite{Heslop:2010kq}.
  Their approach was  different. They
 just assumed a form for the functions that could appear in the answer and then imposed the correct colinear
 limits, which would be the zeroth order  term in the OPE.
  In fact, if one {\it knows} the functions that appear in the answer, then one can find the right combination
  by looking at simple limits. On the other hand, the OPE, can be good for determining, or at least constraining
  the functions that appear. In this case with $R^{1,1}$ kinematics the OPE gives us the functions that appear
  in the answer.

  It is natural to try to extend this two loop analysis to the full $R^{1,3}$ kinematics. In principle, the
  same method works. In practice, it is more complicated because there is a large number of $SL(2)$
   primaries that
  appears. In the case of $R^{1,1}$ kinematics, we had simply one tower of primaries which are  simply plus and
  minus derivatives of
  $ F_{-+}$. In the $R^{1,3}$ case we have primaries of the form $ D_z^l F_{+ z}$
   and $D_{\bar z}^l
  F_{+ \bar z}$, where $z$ and $\bar z$ are the transverse coordinates.
   The $l$ index is  related to powers of one of the extra angular coordinates that we have
  when we do the OPE in the $R^{1,3}$ context. However, we also have primaries of the form   $ D_z^l F_{- z}$ and $D_{\bar z}^l
  F_{- \bar z}$.  When we perform the
   expansion of the one loop answer, we need to separate the contribution from these two sets of primaries.
  We leave this to the future.

    We can also consider higher loop contributions.
    The leading logarithmic terms are easy to obtain, one simply repeats the convolution with the
    one loop anomalous dimension kernel once again.  For $R^{1,1}$ kinematics we wrote an
    expression which reproduced all the $\tau^2$ (and $\sigma^2$) terms in the OPE at three loops. It remains to be seen if
    there are other terms that we should add to that expression in order to reproduce the full answer.
    In principle, we can also get subleading logarithmic terms. These arise from higher loop corrections to the anomalous dimensions, which we can compute, and also from corrections to the OPE coefficients.
    These are harder to compute since they might involve multiparticle states, etc. 
     In short, it would be great to gain analytic control over the breaking of the SL(2) symmetry discussed in the text. This seems to be the main missing link to be able to extend our techniques to any loop order.
      Of course, if one restricts to the lowest terms in the OPE, which only receive contributions from
    single particle exchanges, then it is easy to go to higher loops once one knows the lower loop answers, as explained in
    \cite{OPEpaper}.

    Recently the full $\mathcal{N}=4$ all loop integrand was proposed in a remarkable Yangian invariant form \cite{ArkaniHamed:2010kv}. It would be interesting to see if the OPE limit can be applied directly at this level.
        Also, it was recently understood how to generalize the bosonic null Wilson loops into super loops which are dual to amplitudes with arbitrary polatizations \cite{Mason:2010yk,CaronHuot:2010ek}. It would be very interesting to apply the method we described to compute non MHV scattering amplitudes. In these cases, fermionc excitations of the flux tube will also be excited.
 Another connection which would be interesting to work out concerns the relation between the OPE expansion and the high energy Reggee limit of scattering amplitudes. They seem to be closely related. The recent papers \cite{Bartels:2008ce} seem to be a promising starting point for establishing a more precise connection.

   \section*{Acknowledgments}

We thank F. Alday, N. Arkani-Hamed, B. Basso, F. Cachazo, L.~N.~Lipatov, J.~ Penedones, A.~Prygarin, D. Skinner, D. Volin for very useful discussions. The
work of J.M. was supported in part by the U.S.
Department of Energy grant $\#$DE-FG02-90ER40542. The research of A.S. and P.V. has been supported in part by the Province of Ontario through ERA grant ER 06-02-293. Research at the Perimeter Institute is supported in part by the Government of Canada through NSERC and by the Province of Ontario through MRI.

\appendix

\section{Polygon Wilson loops with $n>10$ sides}
\label{AnyN}

In this Appendix we use the consistency of the OPE expansion to obtain the two loop remainder function for any null polygon in ${R}^{1,1}$ kinematics. As for the octagon and decagon, we start from the $U(1)$ contribution and use the knowledge of the one loop energy (\ref{dispersion}) to obtain the terms linear in $\tau$ in all channels. We will then show that the guess of \cite{Heslop:2010kq} is the simplest solution that is consistent with all the OPE expansion in all channels.

\begin{figure}[t]
\begin{center}
\includegraphics[width=.6\linewidth]{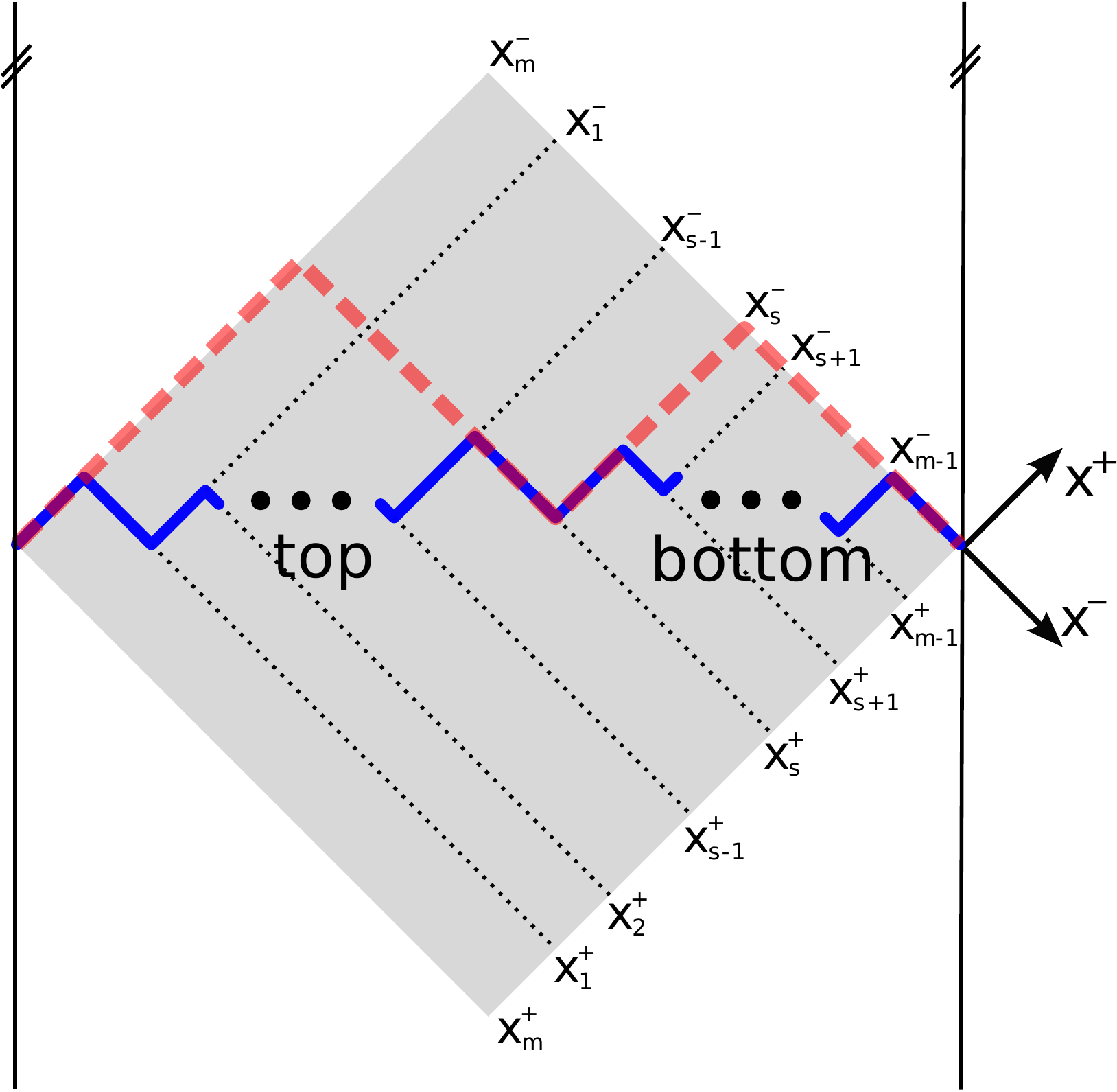}
\end{center}
\caption{Null Wilson loop with $n=2m$ edges in blue embedded in the Penrose diagram of $R^{1,1}$. The red dashed line is the reference square.}\label{GeneralFig}
\end{figure}

Consider a $n=2m$ null polygon in ${R}^{1,1}$. An OPE channel in the $x^+$ direction is defined  by splitting the cusps points $x^+_i$ into two groups of ordered points, {\it top} and {\it bottom}. We denote these by $(x^+_1,x^+_{i+1},\dots,x^+_s)$ and $(x^+_{s+1},\dots,x^+_m)$ respectively, see figure \ref{GeneralFig}. The locations of top and bottom cusps in the $x^-$ direction are $(x^-_m,x^-_1,\dots,x^-_s)$ and $(x^-_s,\dots,x^-_m)$.  We chose the reference square cusps to be at $(x^+_s,x^-_s)$, $(x^+_m,x^-_m)$, $(x_s^+,x_m^-)$ and $(x^+_m,x^-_s)$.

\begin{figure}[t]
\begin{center}
\includegraphics[width=.6\linewidth]{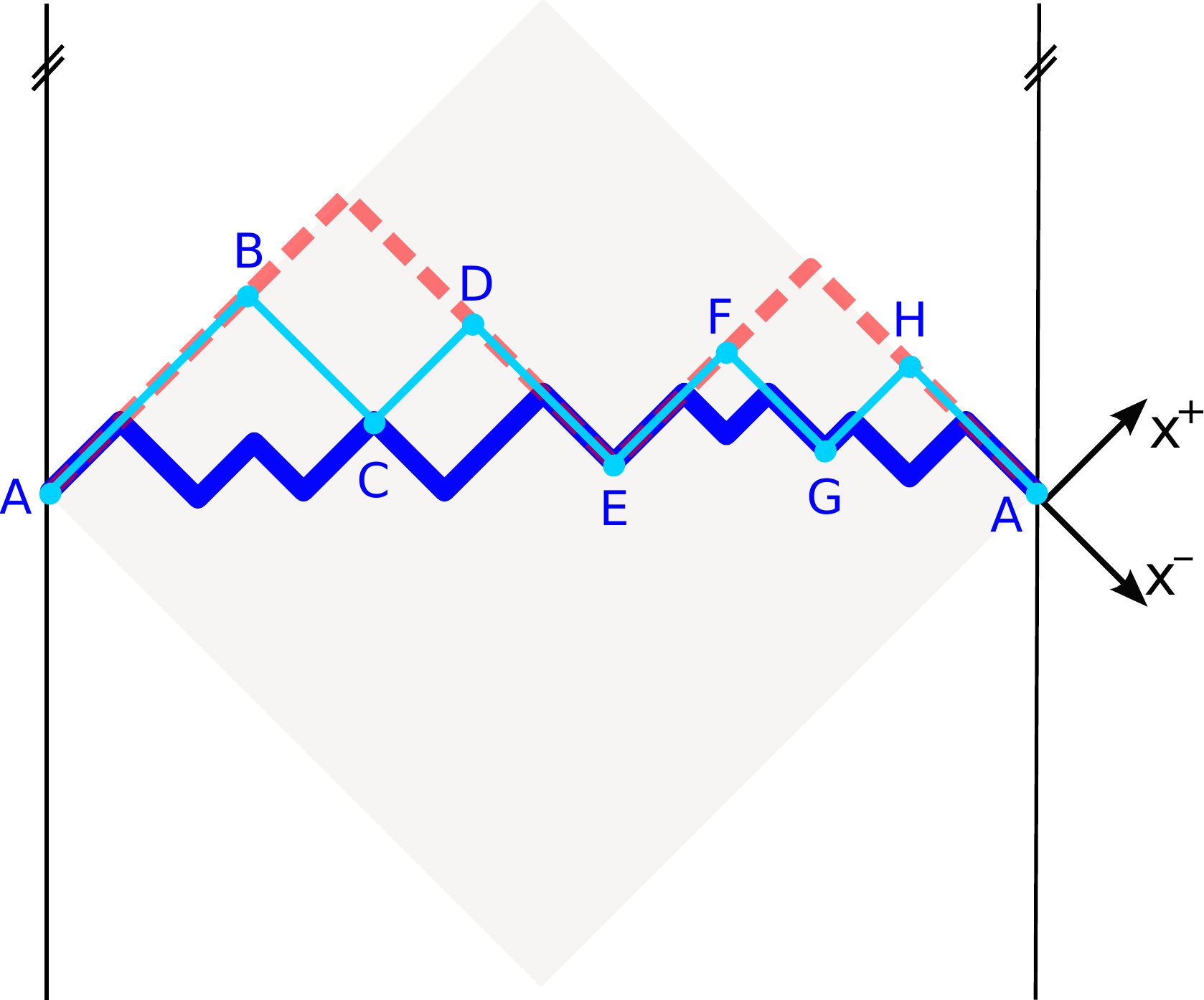}
\end{center}
\caption{Example of a Null Wilson loop (thick dark blue line) together with a reference square (dashed red line) and one of the octagons (thin light blue line). The points $A$ and $E$ are cusps of the reference square. Points $C$ and $G$ are points of the top and bottom part of the polygon respectively. As depicted in this figure, the points $B,D,F,H$ are automatically fixed given these four points.}\label{Example}
\end{figure}

To obtain the $U(1)$ contribution in this channel we decompose the n-gon contour as a sum of octagons as we did at the bottom of figure \ref{DecagonFig}. The cusp of the octagons are denoted by $A,\dots,H$, see figure \ref{Example} for an example.
In this decomposition all octagons share the two cusps of the reference square $A=(x^+_m,x^-_m)$ and $E=(x^+_s,x^-_s)$. Each octagon also shares one more cusp from the top group
\beq
C
\in \{(x_1^+,x_1^-),(x_2^+,x_1^-),(x_2^+,x_2^-),\dots,(x_{s-1}^+,x_{s-2}^-) ,(x_{s-1}^+,x_{s-1}^-) \} \la{topC}
\eeq
and one more cusp from the bottom group
\beq
G
 \in \{(x_{s+1}^+,x_{s+1}^-),(x_{s+2}^+,x_{s+1}^-),\dots,(x_{m-1}^+,x_{m-1}^-) \}
\eeq
So far we described four of the eight cusps of each octagon. It is easy to see that the four remaining cusps are uniquely fixed once these points are chosen, see figure \ref{Example} for illustration. Although we will not explicitly use them in what follows, their positions are
\beq
B=(C^+,A^-) \,\, , \qquad D=(E^+,C^-) \,\, ,  \qquad  F=(G^+,E^-)\,\, ,  \qquad  H=(A^+,G^-)\,.
\eeq
We have therefore a unique decomposition given by a sum over the points $C$ and $G$ (all other points are fixed in terms of these).
The sign of each octagon is determined from its orientation and is given by $(-1)^{(c^+-c^-)+(g^+-g^-)}$ where $c^+$ stands for the index appearing in $C^+$, i.e. $C^+=x^+_{c^+}$ etc.
To summarize, the $U(1)$ part in that channel is given by
\beqa\la{easytau}
R_{U(1)}^{(1,s)}&=&-{g^2\over2}\sum_{C,G}(-1)^{c^+-c^-+g^+-g^-} \log\(1+Y_{GAEC}^+\)\log\(1+Y_{GEAC}^-\)\ .
\eeqa
where $Y_{ABCD}^{\pm}\equiv  \frac{(A^\pm-B^\pm)(C^\pm-D^\pm)}{(A^\pm-C^\pm)(D^\pm-B^\pm)}
$.
Note that $Y_{GAEC}^+ \propto e^{-2\tau}$ and $Y_{GEAC}^-\propto e^{-2\sigma}$.
We can now use (\ref{OPE}) to derive the term in the two loops remainder function that is linear in $\tau$
\beq\la{Generaltau}
R_{\text linear\  in\ \tau}^{(1,s)}=g^4\tau \sum_{C,G}(-1)^{c^+-c^-+g^+-g^-} \log\(1+Y_{GAEC}^+\)\log\(1+Y_{GEAC}^-\)\log\(1+1/Y_{GEAC}^-\)\ .
\eeq
The full two loops remainder function is a function whose term linear in $\tau$ in any channel is given by (\ref{Generaltau}). To guess this function it is useful to notice that the sum over $C$ and $G$ can be partially performed. First we notice that
\beq
1+Y_{GAEC}^+=1+ {(x^+_{g^+}-x^+_m)(x^+_{s}-x^+_{c^+}) \over (x^+_{g^+}-x^+_{s})(x^+_{c^+}-x^+_{m})}= {(x^+_{g^+}-x^+_{c+})(x^+_{s}-x^+_{m}) \over (x^+_{g^+}-x^+_{s})(x^+_{c^+}-x^+_{m})} \equiv {x^+_{g^+,c^+}x^+_{s,m}\over x^+_{g^+,s}x^+_{c^+,m}}
\eeq
Next notice that, from (\ref{topC}), we have
$\sum_{C}  (\dots)= \sum_{c_-}  \[\sum_{c^+=c^-,c^{-}+1} \(\dots \) \]
$
and similar for $G$. We should be slightly more careful with the boundary terms, e.g. when $c^-=s-1$ we only have one allowed value for $c^+$, namely $c^+=s-1$, see (\ref{topC}). In other words, the sums over $c^+$ and $g^+$ are rather trivial as they contain either two or one terms.
E.g.,
\beq
\sum_{c^+=c^-}^{c^-+1}\sum_{g^+=g^-}^{g^-+1} (-1)^{c^+-c^-+g^+-g^-} \log\( {x^+_{g^+,c^+}x^+_{s,m}\over x^+_{g^+,s}x^+_{c^+,m}}\)=\log\( \frac{x^+_{g^-+1,c^-+1} x^+_{g^-,c^-}}{x^+_{g^-+1,c^-} x^+_{g^-+1,c^-}}\) \la{bulk}
\eeq
which is valid away from the boundaries, i.e. for $c^-<m-1$ and $g^-<s-1$. At the boundaries, if $g^-=m-1$ we have
\beq
\sum_{c^+=c^-}^{c^-+1}(-1)^{(c^+-c^-)}\log\( {x^+_{g^+,c^+}x^+_{s,m}\over x^+_{g^+,s}x^+_{c^+,m}}\)=\log\( \frac{x^+_{g^-+1,c^-+1} x^+_{g^-,c^-}}{x^+_{g^-+1,c^-} x^+_{g^-+1,c^-}}\)\la{boundary}
\eeq
with a similar expression for $c^-=s-1$. Results (\ref{bulk}) and (\ref{boundary}) are quite interesting in the sense that they involve the
 logs of  nearest-neighborhood cross-ratios, the indices $s$ and $m$ dropped out completely. These building blocks posses a very useful property:
 $$\log \({x^+_{a+1,b+1}x^+_{a,b}\over x^+_{a+1,b}x^+_{a,b+1}}\)$$
  gives rise to a linear $\tau$ behavior if and only if $a=s$ and $b=m$ or viceversa.  We can use this to write a cyclically symmetric sum which reproduces the $\tau$-linear part in the large $\tau$ limit:
\beqa\la{FinalTwoLoops}
\frac{g^4}{8}\sum_{a,b,c,d \,\mathrm{in\, cyclic\, order}} \log\({x^+_{a+1,c+1}x^+_{a,c}\over x^+_{a+1,c}x^+_{a,c+1}}\)\log\({x^+_{b+1,d+1}x^+_{b,d}\over x^+_{b+1,d}x^+_{b,d+1}}\)  \log\(1+{x^-_{a,d}x^-_{b,c}\over x^-_{a,b}x^-_{c,d}}\)\log\(1+{ x^-_{a,b}x^-_{d,c} \over
x^-_{a,d}x^-_{b,c} }\)\ .\nn
\eeqa
Although it is not manifest, this expression is symmetric in exchanging $x^+$ and $x^-$. Indeed, it is simply the result of a telescopic sum over $x^-$
of the expression guessed in \cite{Heslop:2010kq}, thus showing that it satisfies the $\tau$-linear part of the OPE requirement in all possible channels and is therefore the simplest and most natural solution.

\section{Excitations from Integrability}
\label{IntegrabilityAppendix}

In this appendix we compute the anomalous dimensions of the flux tube excitations using integrability. More precisely we will use the mapping of the flux tube to the GKP state
\beq
 Z \underbrace{ D_+ \dots D_+ }_{S} Z  \,\,\, +\,\,\, \dots \la{without}
\eeq
and study the energy of fluctuations around that state. Here $D_{+}$ is a light-cone direction and $Z$ is one of the complex scalars of the theory which need to be included so that  the derivatives have something to act on. We can think of the scalars as being two fast particles sourcing the flux tube represented by the light-cone derivatives. The spin $S$ is to be taken to infinity as explained shortly. This formula is of course schematic, the state is a complicated superposition as indicated by the dots and an overall trace is omitted.

We are interested in excitations
\beq
 Z D_+ \dots{\chi}  \dots  D_+  Z  \,\,\, +\,\,\, \dots \la{withExcitation}
\eeq
moving in the background of derivatives. The effective length perceived by these excitations is  $2 \log S $. The energy of this state is given by
\beq
E= 2 \gamma_{cusp} \log S + C + \sum_{\text{excitations }\chi} \epsilon_{\chi}(p_{\chi}) \la{energyCusp}
\eeq
where $p_{\chi}$ is the momentum of the excitation and $\epsilon_{\chi}(p)$ is the dispersion relation of the excitation $\chi$. The first term represents the vacuum energy contribution and $C$ is an irrelevant constant, also associated to the vacuum. There are several different excitations we could consider. In this appendix we will consider scalars, fermions and, most relevant for this paper,  the case when
\beq
\chi=D^k_- F_{+-} \,. \la{MainEx}
\eeq
This is the most relevant case since excitations of the form (\ref{MainEx}) are the ones that we expect to generate in the OPE Wilson loop when considering loops in ${R}^{1,1}$ as in this paper. Of course the detailed structure of the Bethe state with excitations (\ref{MainEx}) is quite complicated since it will mix with fermions and scalars. For us what is important is that is has the correct quantum numbers to be created by the polygon Wilson loops. For the bulk of the text, the main result is that these excitations have a one loop energy given by
\beq
\epsilon_{D^k_- F_{+-} }(p)=2+2k+ 2g^2 \( \psi\(1+i\frac{p}{2}\)+ \psi\(1-i\frac{p}{2}\)-2\psi(1) \)+O(g^4) \,. \la{MainEn}
\eeq

Excitations of the GKP string were considered in much greater detail by Benjamin Basso in \cite{BB} who provided all loop expressions for most excitations. In this appendix we will consider mostly one loop and will mainly focus on the excitations needed for the main text. We will also use the method of Baxter polynomials which is very suitable for perturbative weak coupling computations.

\subsection{Scalars $Z$ at one loop}
To construct the state with a single scalar $\chi=Z$ moving in the see of derivatives as in (\ref{withExcitation}) we use Beisert-Staudacher (BS) Bethe equations \cite{BS}. For completeness these equations are written down in section \ref{BS:sec}. The precise notation is introduced in that section.

These equations describe excitations around the $Z^L$ vacuum. From this point of view the excitations are the light-cone derivatives while the vacuum is made out of scalars. In other words, (\ref{withExcitation}) is an extremely excited state. Luckily there is a way of efficiently exchanging particles and vacuum constituents in Integrable models via what is called a particle-hole transformation. This can be nicely implemented using the Baxter polynomial approach as proposed in the $AdS/CFT$ context in \cite{Belitsky:2006en}. We will follow this approach in what follows.

To build (\ref{withExcitation}) we consider BS equations with $S$ momentum carrying excitations (the derivatives) and $L=3$ (the scalars). At one loop the relevant Bethe equations are equivalent to the Baxter TQ relation
\beq
T(u) Q_4(u)= \Phi(u+i/2) Q_4(u+i) + \Phi(u-i/2) Q_4(u-i) \la{Baxter4}
\eeq
where $\Phi(u)=u^L$ with $L=3$,  $Q_4(u)=\prod_{k=1}^S (u-u_j^{(4)})$ and
\beq
T(u)=2\prod_{j=1}^{K_\theta} (u-\theta_j) \,. \la{Tmatrix}
\eeq
The zeros of $T(u)$ are called \textit{holes}.
The index $4$ comes from a conventional labeling of the Bethe roots in the BS equations, see section \ref{BS:sec}. Since $T(u)$ is a polynomial,  the left hand side of (\ref{Baxter4}) vanishes for $u=u_j^{(4)}$ which coincides precisely with Beisert-Staudacher equations when only $Z$'s and $D_+$'s are present in the single trace operator. Note that for (\ref{Baxter4}) to make sense we must have
\beq
K_{\theta}=3 \,.
\eeq
The idea is to trade the dynamics of the $S$ \textit{particles} $u_j^{(4)}$ by the dynamics of the $3$ \textit{holes} $\theta_j$. Actually, as explained in section \ref{large}, two of these holes are located at
\beq
\theta \simeq \pm \frac{S}{\sqrt{2}} \la{largeholes}
\eeq
and do not have therefore any interesting dynamics. They are the so called \textit{large holes}. This result is valid for any coupling, with any finite number of excitations with arbitrary polarizations. The remaining hole is the interesting one, the \textit{small hole}. It corresponds to the excitation $\chi=Z$ moving in the vacuum. The goal is then to find the momentum and energy of this holes, $p_Z(\theta)$ and $\epsilon_Z(\theta)$ to derive $\epsilon_Z(p)$.

At values of $u$ with positive imaginary part the first term in the right hand side of (\ref{Baxter4}) is way bigger than the second one which can therefore be dropped\footnote{This is so for $|u|\ll S$ which is more than enough for our purposes}. The resulting functional equation is then easy to solve \cite{Belitsky:2006en},
\beq
Q_4(u) = c\, 2^{{u}/{i}} \prod_{j=1}^{3} \frac{ \Gamma\(\frac{u-\theta_j}{i}\)}{\Gamma\(\frac{u+i/2}{i}\)} \qquad , \qquad \text{for u in the upper half plane} \,, \la{Qup}
\eeq
where $c$ is a constant which is irrelevant for our argument. For negative imaginary values of $u$ we would drop instead the first term in the right hand side of (\ref{Baxter4}) and we would get the complex conjugate function. We can now compute the energy and momentum of the excitation.

The energy of the state is given by (\ref{EnergyLogQ}). When taking the logarithm of $Q_4$, the product in (\ref{Qup}) becomes a sum of three terms. Two of them are associated to the large holes (\ref{largeholes}) and gives us the vacuum energy contribution, containing a term proportional to $\log S$ plus a constant. The remaining term gives us the energy of the excitation. More precisely, we find (\ref{energyCusp}) with $\gamma_{cusp}=4g^2+\mathcal{O}(g^4)$ and \cite{Belitsky:2006en}
\beq \epsilon_Z(\theta)= 2g^2 (\psi(1/2+i\theta)+\psi(1/2-i\theta)-2 \psi(1) ) \,\,\, +\,\,\, \mathcal{O}(g^4) \,, \la{eZ}
\eeq
The momentum is read off from the effective Bethe equation for the hole excitations. These are nothing but
\beq
\frac{\Phi^+ Q_4^{++}}{\Phi^- Q_4^{--}}=-1 \qquad , \qquad \text{at u=}\theta_j
\eeq
where $f^{\pm}\equiv f(u\pm i/2)$, $f^{++}\equiv f(u+ i)$ etc. This
 is of course the definition of the holes. When plugging (\ref{Qup}) into this equation, the left hand side will contain a factor of the form
 $e^{-i p_Z(\theta)\times 2\log S} $ generated by the large holes in (\ref{Qup}). From this term we read off the momentum $p_Z(\theta)$ (recall that the effective length of the operator is $2\log S$).  In other words,
 $$p_Z(\theta)=\lim_{S\to \infty} \frac{i}{2  \log S}\log \frac{Q_4(\theta+i)}{Q_4(\theta-i)}\,.$$
We find
\beq
p_Z(\theta)=2\theta\,\,\,+\,\,\, \mathcal{O}(g^2) \,. \la{pZ}
\eeq
Combining this result with (\ref{epZ}) we arrive at the known result
\beq
\epsilon_Z(p)= 2g^2 (\psi(1/2+ip/2)+\psi(1/2-ip/2)-2 \psi(1) )\,\,\, +\,\,\, \mathcal{O}(g^4) \,. \la{epZ}
\eeq
\subsection{Other excitations} \la{notation}
To consider other excitations in Beisert-Staudacher equations associated with the Dynkin diagram of figure \ref{Dynkin}, we also need to consider the auxiliary roots $u_j^{(a)}$ for $a \neq 4$. The Baxter equation (\ref{Baxter4}) for the momentum carrying roots is now replaced by
\beq
T Q_4 = \Phi^+ Q_4^{++} Q_3^- Q_5^- + \Phi^- Q_4^{--} Q_3^+ Q_5^+\la{Baxter42}
\eeq
where $T(u)$ is given by (\ref{Tmatrix}) with $K_{\theta}=L+K_3+K_5$. The excitations $u^{(3)}$ and $u^{(5)}$ are fermionic excitations. Note that when $u$ is in the upper half plane the first term in (\ref{Baxter42}) dominates exponentially over the second one. However, when $u \simeq u_j^{(a)}+i/2$ with $a=3,5$  this term is suppressed by $Q_a^-$. This means that, with exponential precision, there is a complex zero of $T(u)$ for each fermionic root located at $u_j^{(a)}+i/2$. Similarly, there is another hole at $u^{(a)}_j-i/2$.

There is an important exception: the case when two roots of type $u^{(3)}_j$ and $u^{(5)}_j$ are in the \textit{same} position $v$. In this case we also have the possibility of having \textit{one} hole at $v+i/2$ and \textit{one} hole at $v-i/2$ instead of \textit{two} holes at $v+i/2$ and two holes  $v-i/2$, as one can conclude after a simple inspection of (\ref{Baxter42}). Hence, from now on, we shall use $v_j$ to denote the roots $u^{(3)}_j=u^{(5)}$ in this particular case, i.e. the which are at the same position and which are accompanied by only two holes at $v_j \pm i/2$. At the same time we shall use $u^{(3)}_j$ and $u^{(5)}_j$ to denote the remaining roots of type $3$ and $5$. Then,
\beqa
T(u)&=&2 \(u^2-\frac{S^2}{2}\)  \prod_{z\in\{v_j\}\cup \{u_j^{(3)}\}\cup \{u_j^{(5)}\}} \(\(u-z\)^2+\frac{1}{4}\) \prod_{z=\{\theta_j\}}(u-z)
\eeqa
where $\theta_j$ are real small holes like the one studied in the previous section which are not bound to any fermionic roots.
Now, as before, we can solve the functional equation (\ref{Baxter42}) by going to the upper half plane and using simple functional identities of Gamma functions. We find
\beq
Q_4(u)= c \, \exp\(-2ui \log S\) \,\frac{1}{\phi^+} \,q_\theta \, q_3^+ \, q_5^+ \, \frac{q_v^+}{q_v^-} \qquad , \qquad \text{for u in the upper half plane} \la{Qup2}
\eeq
where $\phi=\Gamma(u/i)^L$ and
\beq
q_a\equiv \prod_{j=1}^{N_{a}} \Gamma\(\frac{u-u_j^{(a)}}{i}\)\,.
\eeq
Note that $Q_4$ vanishes for $u=v+i/2$ which means that there is a complex root $u_4$ at this position. Similarly from the expression for $Q_4(u)$ in the lower half plane we conclude that there is another root $u_4$ at $v-i/2$. Figure \ref{excitations} summarizes the three possibilities we found so far: single real hole, fermionic excitation dressed by two complex holes and double fermionic composite dressed by two complex holes and two complex momentum carrying roots. In this picture we only represent the holes and the particles $u^{(a)}$ with $a=3,4,5$. All other auxiliary roots don't couple directly to the momentum carrying roots $u^{(4)}$ and hence carry no energy nor momentum at this loop order. When considering two loops and higher we would need to consider the energy of excitations $u^{(1)}$ and $u^{(7)}$ as well.

\begin{figure}[t]
    \centering
       \resizebox{80mm}{!}{\includegraphics{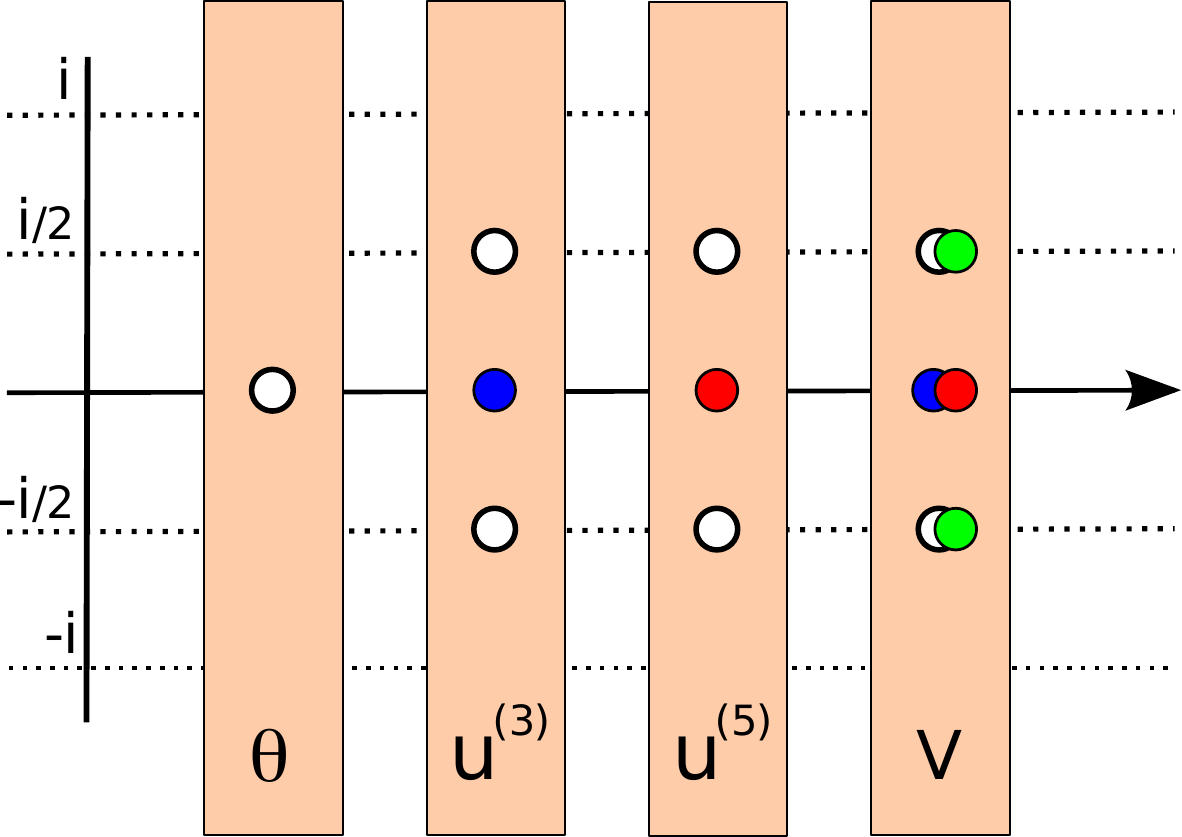}}
        \caption{\small Fundamental building blocks using the holes and the fermionic roots $u^{(3)}$ and $u^{(5)}$. White balls represent holes; Blue balls correspond to fermionic roots $u^{(3)}$; Red balls indicate fermionic roots $u^{(5)}$; Green balls denote complex momentum carrying roots $u^{(4)}$. The energy of the first building block (the single hole) is given by (\ref{epZ}) while the energy of the fermionic excitations in the middle is given by (\ref{epsip}). The last composite, to the right, is isotopic to all loop orders \cite{BB} and hence carries no energy or momentum.
         \la{excitations}}
\end{figure}

We can now use again (\ref{EnergyLogQ}) to read of the energy in the presence of all possible excitations. We find
\beq
E- E_{vacuum}=\sum_{z\in \{u_j^{(3)}\} \cup \{u_j^{(5)}\}} \epsilon_{\psi}(z)+\sum_{z\in \{\theta_j\}}  \epsilon_Z(z) \la{pol}\,,
\eeq
where (\ref{epZ}) and the energy of $u^{(3)}$ and $u^{(5)}$ excitations are given by
\beq
\epsilon_{\psi}(\theta)=2g^2\(\psi(1+i\theta)+\psi(1-i\theta)-2\psi(1)\) \,\,\, +\,\,\, \mathcal{O}(g^4) \,. \la{epsi}
\eeq
Note that the excitations $v_j$ carry no energy at all!, they are not present in (\ref{pol}). We can now compute the momentum of the several excitations. The momentum is read off from the term $e^{i p(\theta) 2 \log S}$ arising in the corresponding Bethe equation in (\ref{BSeqs}). For composite states we should multiply the corresponding equations as usual. Plugging the corresponding $Q_4$ given by (\ref{Qup2}) in (\ref{BSeqs}) we read the momentum of these fluctuations as before. For $u^{(3)}$ and $u^{(5)}$ excitations we find again
\beq
p_{\psi}(\theta)=2\theta \,\,\,+\,\,\, \mathcal{O}(g^2)\la{ppsi}
\eeq
while the $v$ excitations have carry no momentum,
\beq
p_v(\theta)=0 \,\,\,+\,\,\, \mathcal{O}(g^2)\,.
\eeq
In sum we found
\beq
\epsilon_{\psi}(p)=2g^2\[\psi(1+ip/2)+\psi(1-ip/2)-2\psi(1)\] \la{epsip}\,.
\eeq
Using these results we can derive the dispersion relation of several different excitations. First, consider the excitation associated with a single root $u^{(3)}$. An excitation of this type carries the quantum numbers of $c_4 b_1$, see figure \ref{Dynkin}. This bilinear converts
\beq
D_{+} Z \leftrightarrow D_{1\dot 1} \Phi_{43} \leftrightarrow b_1^{\dagger} a_{\dot 1}^{\dagger} c_4^{\dagger} c_3^{\dagger} |0\rangle \la{DZ}
\eeq
into a twist one fermionic excitation
$$
\Psi_{\dot 1 3} \leftrightarrow  a_{\dot 1}^{\dagger} c_3^{\dagger} |0\rangle
$$
Hence we derived that
\beq
\epsilon_{\psi \, \text{twist one}}(p)=2g^2\[\psi(1+ip/2)+\psi(1-ip/2)-2\psi(1)\]\,.
\eeq
Suppose we added a bosonic root $u^{(2)}$ to this excitation. This would amount to acting with the oscillators $c_3 c^\dagger_4$ which would convert the R-charge index $3$ of the fermion into a $4$. This would be another fermionic twist one excitation. Since $u^{(2)}$ excitations do not carry energy or momentum this excitation would have the same anomalous dimension which is of course what we expect.

Lets now consider a real hole $\theta$ together with a composite of type $v$ described above. As we saw the $v$ composite carries no momentum or energy at all. There is a simple reason \cite{BB}\footnote{We thank Benjamin Basso for illuminating discussions on this point. See \cite{BB} for more details.} for this which becomes clear when we identify the excitation using the oscillator picture. By adding one root $u^{(3)}$ and one root $u^{(5)}$ we convert the scalar excitation $Z=\Phi_{43}$ into another scalar excitation $Y=\Phi_{32}$. The $v$ composite is an isotopic degree of freedom responsible for implementing the $SO(6)$ symmetry of the GKP vacuum! This is why it has zero energy and momentum as we checked at one loop order. Of course, from this symmetry argument the $v$ composite ought to have zero energy and momentum at any loop order and it does indeed \cite{BB}.

\begin{figure}[t]
    \centering
       \resizebox{150mm}{!}{\includegraphics{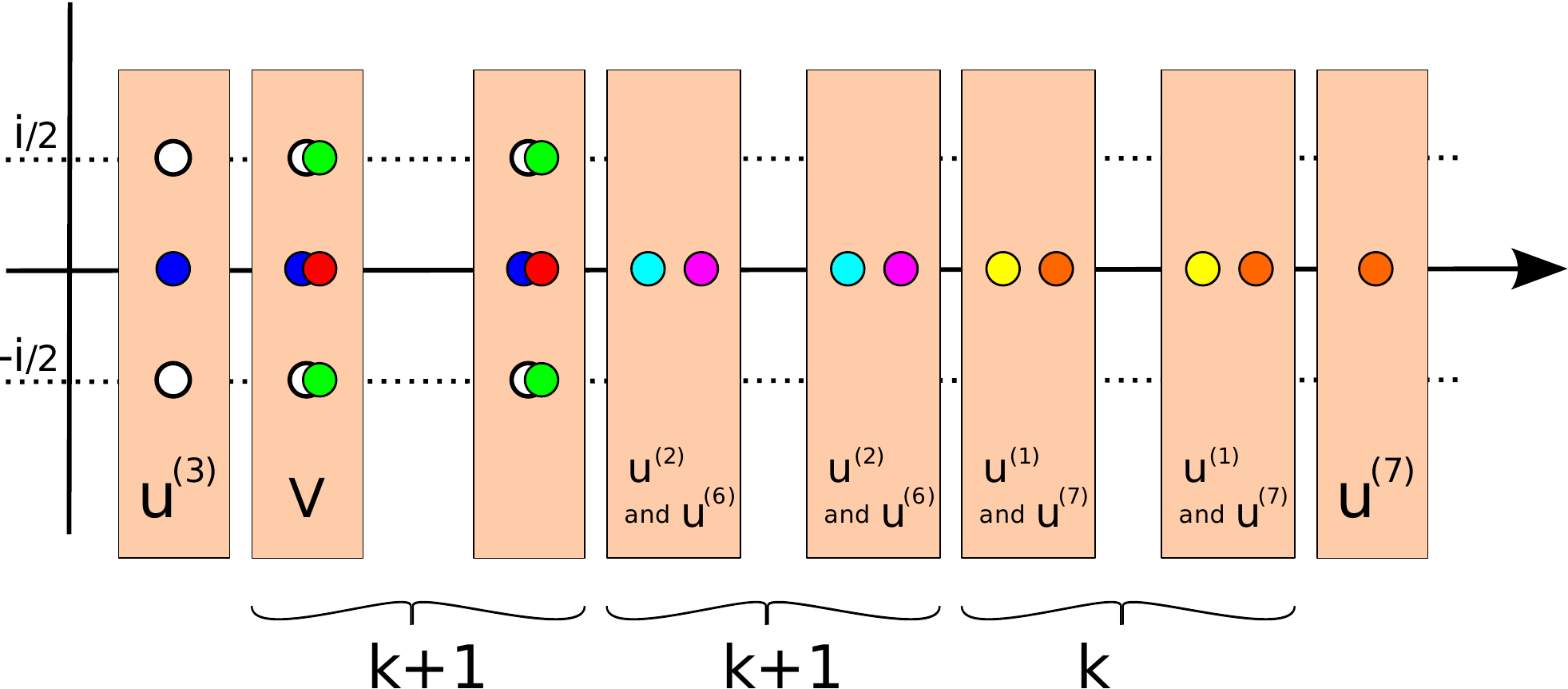}}
        \caption{\small Pattern of Bethe roots corresponding to the excitation $D_-^k \dot{\mathcal F}_{\dot 1\dot2}$. This pattern of roots is what we would get if we would add small twists to the Bethe equations to regulate them. Otherwise we would find instead a descendent of $D_-^k \dot{\mathcal F}_{\dot 1\dot2}$. In other words, some of the roots in this figure would go to infinity. The excitation $D_-^k{\mathcal F}_{12}$ is trivially related to this one by a wing exchange $u_j^{(a)}\leftrightarrow u_j^{(8-a)}$. It has therefore the exact same energy and momentum. The state (\ref{MainEx}) is nothing but the linear combination $D_-^k{\mathcal F}_{12}+D_-^k\dot {\mathcal F}_{\dot 1\dot 2}$ and therefore it has the same energy as either of these states. Now, at one loop, only the first constituent in this figure -- the single root $u^{(3)}$ -- carries energy and momentum! All other excitations are isotopic. Hence (\ref{MainEn}) follows.
         \la{pyr}}
\end{figure}
Finally we could consider more complicateds excitation such as (\ref{MainEx}). It is described by several roots of all possible kinds, see figure \ref{pyr}. However, we see that only a single fermionic root carries energy at one loop order. Hence (\ref{MainEn}) follows. At higher loops the roots $u_j^{(1)}$ and $u_j^{(7)}$ are no longer isotopic in this vacuum. Thus the state (\ref{MainEx}) no longer has a reasonable description in terms of a single particle state; instead it decays into several excitations (e.g. fermions).

Now, this result is part of a much more general observation which is that at one loop the roots $u^{(1)}$ and $u^{(7)}$ are isotopic and hence there is an isotopic symmetry enhancement which will be broken at higher loops when these particles become momentum and energy carrying.  This symmetry contains the $SL(2,R)$ symmetry mentioned in the main text. It implies that the one loop spectrum is organized according to $SL(2,R)$ primaries and descendents. For example, all excitations $D_-^k \psi$ (or $D_-^k X$) have the same anomalous dimension because the pattern of Bethe roots associated to this excitation is just the Bethe roots of $\psi$ (or $X$) plus a bunch of isotopic roots as in figure \ref{pyr}.  At higher loops the roots $u^{(1)}$ and $u^{(7)}$ start carrying energy and momentum and all these degeneracies are lifted.

\subsubsection*{One loop Summary}
We have found that the one loop dispersion relation for excitations around the flux tube is always of the form
\beq
\epsilon_{s}(p)=2g^2\[\psi(s/2+ip/2)+\psi(s/2-ip/2)-2\psi(1)\] \la{epsips}\,, \nn
\eeq
where $s=S+\Delta$ is the conformal spin of the excitation. For example
\beqa
\begin{array}{lll}
s&&\text{excitations}  \\ \hline
1&:&F_{-i}\ \ \Phi_{ab}\ \ D_-\Phi_{ab}\ \ \psi_{2a}\ \ \psi_{\dot 2 a}     \\
2&:&F_{+-}\ F_{12}\ \ \, D_i\Phi_{ab}\ \ \ \psi_{1a}\ \ \psi_{\dot 1 a}      \\
3&:&F_{+i}  
\end{array}\ . \nn
\eeqa

\subsection{Higher loops}
In this section we review how to set up the computation of the anomalous dimensions at higher loop orders. We will illustrate how to perform this computation at $2$ loop order but the generalization to higher loops can be done along the same lines. For all loop results see \cite{BB}.
Up to four loops the Baxter equation reads
\beq \la{baxterr}
T Q_4 =\frac{ \Phi^+ Q_4^{++} Q_3^- Q_5^- }{(B_4^{(+)+})^2 B_3^{+} B_5^{+} B_1^+ B_7^+} + c.c.
\eeq
See section \ref{BS:sec} for the definition of the several Baxter polynomials.
Up to wrapping order $g^{2L}$, the right hand side of this equation is a polynomial in $u$ which allows us to define a polynomial $T(u)$ thus simplifying the computations. This justifies the splitting of terms between the displayed terms and those in "c.c.". Now we proceed as before. For a very similar computation see \cite{Belitsky:2006wg}.
For positive imaginary part of $u$ we can drop the "c.c." terms in (\ref{baxterr}) and find
\beq\la{Qcorr}
\frac{Q_4(u)}{ Q^{(0)}_4(u)}=1+g^2 i\[L\,\d_u+2  \partial_u \log Q_4^{(0)} \Big|_{u=i/2}+ \sum_{a \, \text{odd}} \partial_u \log Q_a\Big|_{u=0}\]\psi({\textstyle {1\over2}}-iu)+\mathcal{O}(g^4)\, ,
\eeq
where $Q_4^{(0)}(u)$ is given by (\ref{Qup2}).
As before, the energy of the several excitations can then be extracted from (\ref{EnergyLogQ}) while their momentum is identified by looking at the term proportional to $2\log S$ in the (logarithm of) the corresponding Bethe equation. For holes and fermionic excitations of type $u^{(3)}$ and $u^{(5)}$ the formulae of the previous section are corrected to
\beqa\la{energy2loop}
\epsilon_{Z}(u)&=&+\gamma_{cusp}\,\, \psi_0^{(+)}(1/2+iu)\,-\,\,2g^4\, \psi_2^{(+)}(1/2+iu)  \,, \\
\epsilon_{\psi}(u)&=& +\gamma_{cusp}\,\, \psi^{(+)}_0(1+iu)\,\,\,\,\,\,\,-\,\,2g^4\, \psi _2^{(+)}(1+i u) \,, \nn
\eeqa
where
\beq
\psi^{(+)}_{a}(b+iu)= {1\over2}\[\psi_a(b+iu)+\psi_a(b-iu)-2(a+1)\psi_a(1)\]
\eeq
Roots of type $1$ and $7$ only acquire energy at $3$ loops, at $2$ loops we still have $\epsilon_{1/7}(u)=0$.
For the momentum we find
\beqa\la{momentum2loop}
p_{Z}(u)&=&2u-\,2\pi g^2\tanh(\pi u)\,,\\
p_{\psi}(u)&=& 2u-\,2\pi g^2\coth(\pi u)\,,\nn\\
p_{1/7}(u)&=&\,\,\,\,\,\,  \ +\,\,2g^2/u\nn\,.
\eeqa

\subsection{The two large holes} \la{large}
In this appendix we explain why there are two (and only two) large holes located at $\pm S/\sqrt{2}$.
The argument relies on two simple observations.
\begin{enumerate}
\item Consider
\beq
\frac{T(u)}{Q_3(u)Q_5(u) \Phi(u)} \equiv 2\cos p(u)
\eeq
at large values of $u \sim S$. For example, at one loop we have, see (\ref{Baxter42}),
\beq
p(u)=\frac{L}{2u}+\sum_{j=1}^S \frac{1}{u-u_j^{(4)}} -\frac{1}{2} \sum_{j=1}^{K_3} \frac{1}{u-u_j^{(3)}}-\frac{1}{2} \sum_{j=1}^{K_5} \frac{1}{u-u_j^{(5)}}
\eeq
The precise form of this expression is not very relevant (e.g. at higher loops this expression is modified but its form as a sum of poles still holds and the argument that follows goes through untouched).

For $u\simeq S$ the roots are well described by a density and the function $p(u)$ develops two large cuts (one for positive $u$ and another for negative $u$). The real part of $p(u)$ on these curs coincides with the  (logarithm of the) Bethe equations in the scaling limit and is therefore equal to the mode number of the corresponding roots. For the GKP ground state, it equals $\pi$ ($-\pi$) for positive (negative) $u$.  For even larger $u$'s, i.e. after the cut, the function $p(u)$ decays to zero. Between $p(u)=\pi$ and $p(u)\to 0$ there is only one value where $\cos p(u) $ vanishes which is of course $p(u)=\pi/2$. Thus there is only one large hole for positive $u$. Similarly there is only one large hole for negative $u$.
\item At large $u$ we have
\beq
T(u)= u^{L+K_3+K_5} \(2+ \frac{t_2}{u^2}+\frac{t_4}{u^4}+\dots\)
\eeq
where
\beq
t_2=S^2+\mathcal{O}(S) \,.
 \eeq
 This follows trivially from expanding (\ref{Baxter42}) at large $u$. Now, the existence of only two large holes proved above implies that $t_{2k}/u^{2k}\ll t_2/u^2$ for $k>1$. Thus, for large $u$, we have $T(u)\propto 2u^2-S^2$ so that the two large holes, which are the two large zeros of $T(u)$, are indeed located at $\pm S/\sqrt{2}$.\footnote{It is also fun to note that at the end of the cut we have $p(u)=\pi$ which means $2+ \frac{t_2}{u^2} \simeq -2$ yielding $u\simeq S/2$ which is indeed the known value for the end of the large cuts. }
\end{enumerate}

\subsection{The $\mathcal{N}=4$ Asymptotic Bethe Ansatz} \la{BS:sec}
In $\mathcal{N}=4$ SYM single trace \textit{words} are made out of \textit{letters} which can be written using bilinears of oscillators such as
\beq
\Phi_{ab} \leftrightarrow c^{\dagger}_a c^{\dagger}_b |0\rangle \,\, , \,\,\, \mathcal{D}_{\dot \alpha \beta}\leftrightarrow a^{\dag}_{\dot \alpha} b^{\dag}_{\beta} |0\rangle\,\, , \,\,\,\mathcal{F}_{\alpha \beta}\leftrightarrow b^{\dag}_{\alpha}b^{\dag}_{\beta}|0\rangle
\,\, , \,\,\, \dot{\mathcal{F}}_{\dot\alpha \dot\beta}\leftrightarrow a^{\dag}_{\dot\alpha}a^{\dag}_{\dot\beta}|0\rangle \,\, , \,\,\,
\Psi_{a\alpha} \leftrightarrow c^{\dagger}_a b^{\dagger}_{\alpha} |0\rangle \,\,, \text{etc} \,.\nn
\eeq
The oscillators can have $SU(2)\times SU(2)$ Lorentz indices ($a^{\dagger}_{\dot \alpha}$ and $b^{\dagger}_{\alpha}$) or $SU(4)$ $R$-charge index ($c^{\dagger}_a$). The oscillators $a$ and $b$ are bosonic while $c$ is fermionic.   Difference single traces are made out of different numbers of bilinears, denoted by $K_a$, see figure \ref{Dynkin} and \cite{BS, BS2, Beisert:2004ry} for more details. The anomalous dimensions of states for fixed $K_a$'s are then given by solving $K_1+\dots K_7$ algebraic equations for the so called Bethe roots $\{u_j^{(a)}\}$ where $a=1,\dots,7$ and $j=1,\dots , K_a$. For example the GKP vacuum is made out of $S$ light-cone derivatives $D_+=D_{1\dot1}$, i.e. we we can study this state by considering $S$ excitations of type $u^{(a)}_j$. Once the Bethe roots are found we can read of the anomalous dimension of the corresponding single trace through
\beq
E=2 i g^2  \(\log Q_4 \)'(i/2) + i g^4 \(\log Q_4 \)'''(i/2) + c.c. + \mathcal{O}(g^6) \la{EnergyLogQ}
\eeq
where $$Q_a(u)\equiv \prod_{j=1}^{K_{a}} (u-u_{j}^{(a)})\,.$$ The algebraic equations are the Beisert-Staudacher equations
\begin{figure}[t]
    \centering
       \resizebox{120mm}{!}{\includegraphics{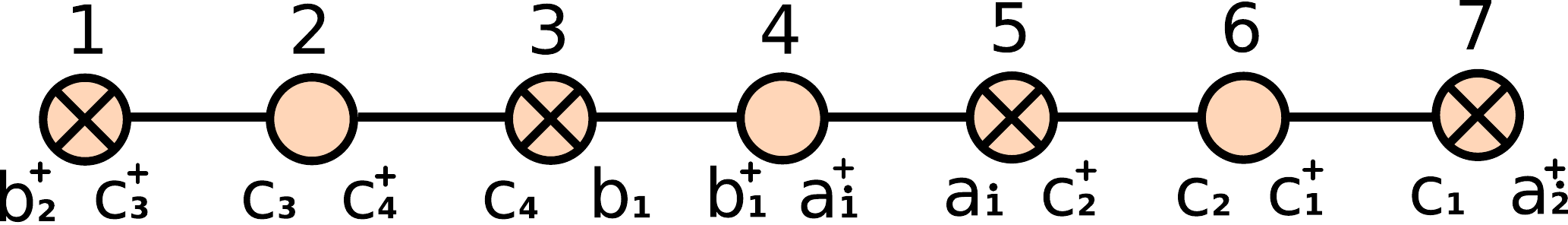}}
        \caption{\small Beisert-Staudacher Dynkin diagram and corresponding oscillators.
         \la{Dynkin}}
\end{figure}
\beq\la{BSeqs}
\begin{array}{lllllllllllllll}
+1  &=& & &  \ds\frac{Q_2^{-} }{Q_2^{+}} & & \ds  \color{blue}\qquad\,\,\,\,\,\frac{B_4^{(-)} }{B_4^{(+)}} & & & & & \text{at} & u=u_j^{(1)}\\
-1 &=& & \ds \frac{Q_1^{-} }{Q_1^{+}} & \ds \frac{Q_2^{++} }{Q_2^{--}}  & \ds\frac{Q_3^{-} }{Q_3^{+}} & & & & & &    \text{at} & u=u_j^{(2)}\\
+1 &=& & & \ds\frac{Q_2^{-} }{Q_2^{+}} & & \ds\frac{Q_4^{+} }{Q_4^{-}} \color{blue} \,\,\,\,\,\,\,\,\frac{B_4^{(-)} }{B_4^{(+)}} & & & & &    \text{at} & u=u_j^{(3)}\\
-1 &=&\ds\frac{\Phi^{-}}{\Phi^{+}}  & \color{blue} \ds \frac{B_1^{+} }{B_1^{-}}   & &\ds \color{black} \frac{Q_3^{+} }{Q_3^{-}}\color{blue} \frac{B_3^{+} }{B_3^{-}}  & \ds \frac{Q_4^{--} }{Q_4^{++}} \color{blue} \[\frac{B_4^{(+)+} }{B_4^{(-)-}} \sigma \]^2&\ds \frac{Q_5^{+} }{Q_5^{-}}\color{blue} \frac{B_5^{+} }{B_5^{-}}  & & \color{blue}\ds \frac{B_7^{+} }{B_7^{-}}& &    \text{at} & u=u_j^{(4)}\\
+1 & = & & & & & \ds \frac{Q_4^{+} }{Q_4^{-}} \,\,\,\,\,\,\,\,\color{blue} \frac{B_4^{(-)} }{B_4^{(+)}} & & \ds \frac{Q_6^{-} }{Q_6^{+}} & & & \text{at} & u=u_j^{(5)} \\
-1 & = & & & & & &  \ds \frac{Q_5^{-} }{Q_5^{+}}&\ds \frac{Q_6^{++} }{Q_6^{--}} & \ds \frac{Q_7^{-} }{Q_7^{+}}  & & \text{at} &u=u_j^{(6)} \\
+1 &= & & & & & \color{blue}\ds\qquad \,\,\,\,\,\frac{B_4^{(-)} }{B_4^{(+)}}  &     & \ds \frac{Q_6^{-} }{Q_6^{+}} & & & \text{at} & u=u_j^{(7)}
\end{array}
\eeq
Where the form of $\Phi(u)=u^L+ \mathcal{O}(g^2) $ and $\sigma^2(u)=1+\mathcal{O}(g^4) $ is irrelevant for us while
\beq
B_a(u)=1+\frac{g^2}{u}\sum_{j=1}^{K_a} \frac{1}{u_j^{(a)}} +  \mathcal{O}(g^4) \qquad,\qquad
B_a^{(\pm)}(u)=1+\frac{g^2}{u}\sum_{j=1}^{K_a} \frac{1}{u_j^{(a)}\mp i/2}+ \mathcal{O}(g^4) \,.
\eeq
In (\ref{BSeqs}) we colored blue the terms which can be set to one at one loop. In this case the Bethe equations take the usual universal rational form
\beq
\(\frac{u_j^{(a)}+\frac{i}{2} \delta_{a4}}{u_j^{(a)}-\frac{i}{2} \delta_{a4}}\)^L=\prod_{k,b \neq j,a} \frac{u_j^{(a)}-u_k^{(b)}+i M_{ab}}{u_j^{(a)}-u_k^{(b)}-i M_{ab}}
\eeq
where $M_{ab}$ is (one of the possible forms of) the $PSU(2,2|4)$ Dynkin matrix \cite{BS2}.
\begin{figure}[t]
    \centering
       \resizebox{130mm}{!}{\includegraphics{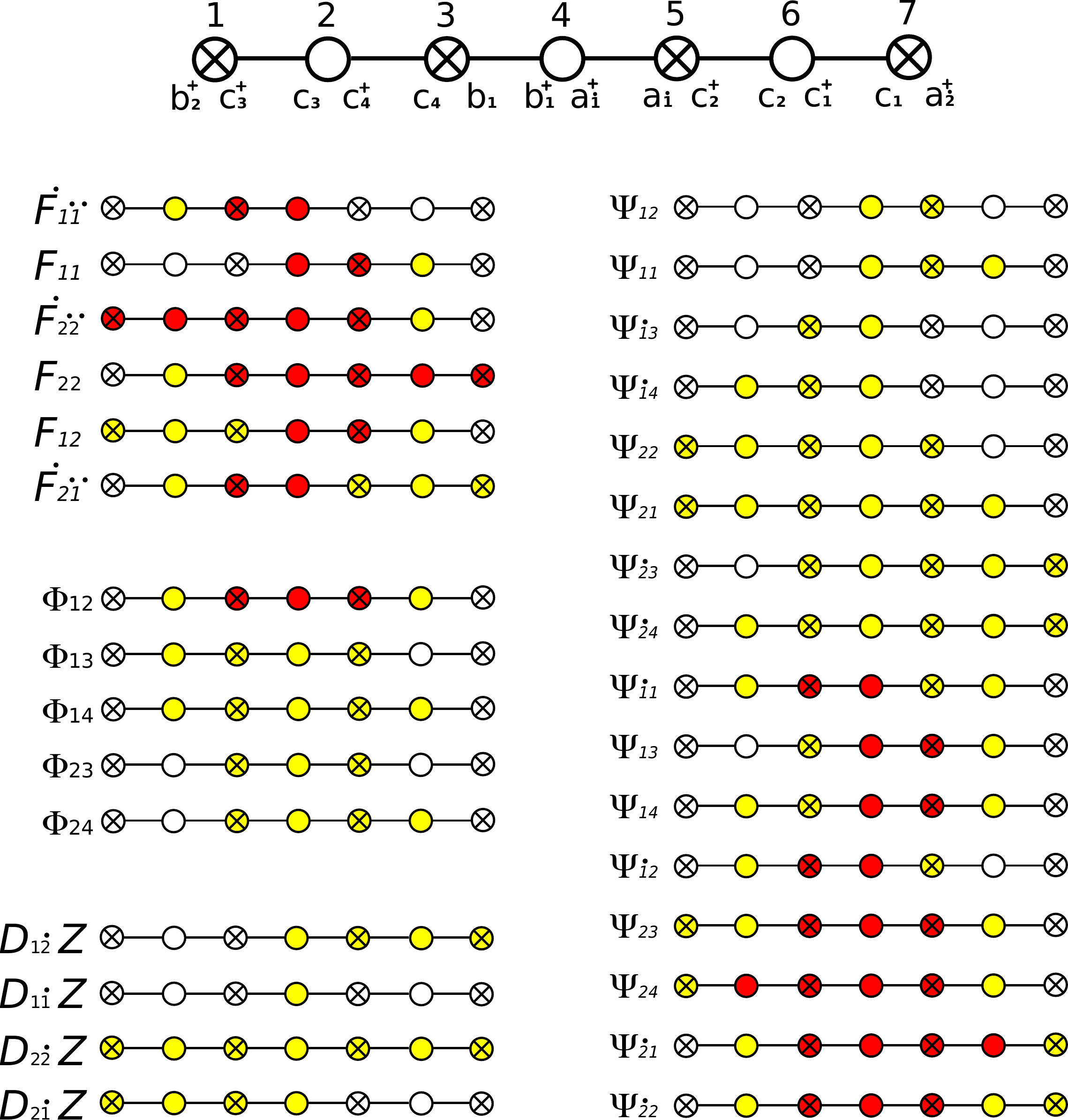}}
        \caption{\small The several excitations around the BMN vacuum.
         \la{excitationsTwo}}
\end{figure}

Finally, let us end this section with a short useful dictionary between Lorents and $SU(2)\times SU(2)$ indices, see e.g.  \cite{Beisert:2004ry}. We have
\beqa
\begin{array}{lll}
D_{1\dot1}&= D_{0+3}&=D_+ \\
D_{2\dot2}&= D_{0-3}&=D_- \\
D_{2\dot1}&= D_{1+i 2}&=D_{\bf 1} \\
D_{1\dot2}&= D_{1-i 2}&=D_{\bf 2}
\end{array} \qquad , \qquad
\begin{array}{ll}
F_{+ \bf 1}&= -2\, \mathcal{F}_{22} \\
F_{+ \bf 2}&= -2\, \dot{\mathcal{F}}_{\dot 2\dot2}  \\
F_{- \bf 1}&= +2 \,\mathcal{F}_{11} \\
F_{- \bf 2}&= +2\, \dot{\mathcal{F}}_{\dot 1\dot1}  \\
F_{+ -}&= -\mathcal{F}_{12}-\dot{\mathcal{F}}_{\dot 1\dot 2} \\
F_{ \bf 1 2}&=  -\mathcal{F}_{12}+\dot{\mathcal{F}}_{\dot 1\dot 2}
\end{array}
\eeqa

\end{document}